\documentclass[12pt]{article}
\usepackage{a4wide}
\usepackage{axodraw}
\usepackage{floatflt}
\usepackage{amssymb}
\usepackage{graphicx}
\newcommand{\dsp}{\displaystyle}
\newcommand{\ba}{\begin{eqnarray}}
\newcommand{\ea}{\end{eqnarray}}
\newcommand{\be}{\begin{equation}}
\newcommand{\ee}{\end{equation}}
\newcommand{\re}{\mbox{Re}}
\newcommand{\im}{\mbox{Im}}
\newcommand{\kob}{\ensuremath{\overline{K^0}}}
\newcommand{\hepph}[1]{{\tt hep-ph/#1}}
\begin{document}
\begin{titlepage}
\begin{flushright}
LU TP 00-44\\
hep-ph/0010265\\
October 2000
\end{flushright}
\vfill
\begin{center}
{\large\bf WEAK INTERACTIONS OF LIGHT FLAVOURS
\footnote{Lectures given at the Advanced School on QCD at Benasque, Spain,
3-6 July 2000\\
Partially supported by the European Union TMR Network
EURODAPHNE (Contract No. ERBFMX-CT98-0169).}}\\[2cm]
{\large\bf Johan Bijnens}\\[0.5cm]
Dept. of Theoretical Physics, Lund University,\\
S\"olvegatan 14A, S22362 Lund, Sweden
\end{center}
\vfill
\begin{abstract}
An overview is given of weak interaction physics of the light flavours.
It starts with the definition of the CKM matrix and the measurement of
its components in the light-flavour sector via semi-leptonic decays.
The main part of the lectures is devoted to non-leptonic decays with a main
emphasis on analytical calculations of $K\to\pi\pi$ and $K\leftrightarrow
\kob$. It finishes with an overview of Chiral Perturbation Theory
in $K\to3\pi$ and of rare Kaon decays.
\end{abstract}
\vfill

\end{titlepage}

%\tableofcontents

\section{Introduction}

These lectures on the weak interaction are not out of place at a school
devoted to the strong interaction. As will be extremely obvious in the
remainder the main problem in dealing with the weak interactions of quarks
is that they are confined in hadrons and thus the non-perturbative regime
of the strong interactions becomes very important.

The lectures first present an introduction as to why we want to study
this type of physics and the Cabibbo-Kobayashi-Maskawa matrix.
The second part then discusses semi-leptonic decays. The relevant 
physics here is the 
determination of $|V_{ud}|$ and $|V_{us}|$.
This will allow us to 
test precisely the unitarity relation
\be
|V_{ud}|^2+|V_{us}|^2+|V_{ub}|^2=1\,.
\ee

The main part of the lectures is dedicated to the calculation of
non-leptonic decays and in particular of $K\to\pi\pi$ and
$K$-$\kob$ mixing. This 
tests the $CP$-violating part of the standard model and in particular
the consistency
of the whole $CKM$-ansatz in the Higgs-Fermion part of the standard model.
Here Kaon physics is very complementary to $B$-physics.
Loop diagrams and QCD effects are {\em very} important.

We conclude by a short overview of
applications of Chiral Perturbation Theory (CHPT) to $K\to3\pi$ and
of rare Kaon decays, those which are first tests of
 QCD and the Chiral Lagrangian and those
important for the SM (and beyond) weak interaction part.

The lectures by Ben Grinstein \cite{Grinstein}
cover the complementary aspects in
$B$-physics. Gerhard Ecker \cite{Ecker}
described Chiral Perturbation Theory (CHPT).
Those techniques also play an important role here. Finally, Chris Sachrajda
\cite{Sachrajda}
described the present status of the lattice QCD calculations of basically
the same quantities.

Some other lecture notes or review articles
covering similar topics are Refs. \cite{Buras1} to \cite{Isidori}.

\section{Standard Model}

The Standard Model Lagrangian has four parts:
$$
{\cal L}_{SM} =
\dsp\underbrace{{\cal L}_H(\phi)}_{\mbox{Higgs}}
+
\underbrace{{\cal L}_G(W,Z,G)}_{\mbox{Gauge}}
\dsp+\underbrace{\sum_{\psi=\mbox{fermions}}
\bar\psi iD\hskip-0.7em/\hskip0.4em \psi}_{\mbox{gauge-fermion}}
+
\underbrace{\sum_{\psi,\psi^\prime=\mbox{fermions}}
g_{\psi\psi'}\bar\psi\phi\psi'}_{\mbox{Yukawa}}
$$
The experimental tests on the various parts are at a
very different level:\\[0.5cm]
\begin{tabular}{ll}
gauge-fermion& Very well tested at LEP-1 and other precision measurements\\
Higgs& Hardly tested, real tests coming up now at LEP-2, Tevatron
and\\& LHC in the future\\
Gauge part& Well tested in QCD, partly in electroweak at LEP-2\\
Yukawa& The real testing ground for weak interactions and the main source
of\\& the number of standard model parameters.
\end{tabular}
\vskip0.5cm

There are three discrete symmetries that play a special role, they are\\
$\bullet$ C Charge Conjugation\\
$\bullet$ P Parity\\
$\bullet$ T Time Reversal\\[0.2cm]
QCD and QED conserve C,P,T separately. Local Field theory by itself
implies CPT. The fermion and Higgs\footnote{This is only true for the
simplest Higgs sector. In the case of two or more Higgs doublets
a complex phase can appear in the ratios of the vacuum expectation values.
This is known as spontaneous CP-violation or Weinberg's mechanism.}
part of the SM-lagrangian conserves CP and T
as well.

The only part that violates CP and as a consequence also T is
the Yukawa part.
The Higgs part is responsible for two parameters, the gauge part for three
and the Higgs-Fermion part contains in principle 27 complex parameters,
neglecting Yukawa couplings to neutrinos.
Luckily most of the 54 real parameters in the Yukawa sector are unobservable.
After diagonalizing the lepton sector there only the three charged lepton
masses remain. The quark sector can be similarly diagonalized
leading to 6 quark masses, but some parts remain in the difference between
weak interaction eigenstates and mass-eigenstates. The latter is conventionally
put in the couplings of the charged $W$-boson, which is given by
\ba
&\dsp
-\frac{g}{2\sqrt{2}}W^-_\mu 
\left(\overline u^\alpha~\overline c^\alpha~\overline t^\alpha\right)
\gamma^\mu\left(1-\gamma_5\right)
\left(\begin{array}{ccc}V_{ud}&V_{us}&V_{ub}\\
                        V_{cd}&V_{cs}&V_{cb}\\
                        V_{td}&V_{ts}&V_{td}\end{array}\right)
\left(\begin{array}{c}d_\alpha\\s_\alpha\\b_\alpha\end{array}\right)
&\nonumber\\ \nonumber&
\dsp-\frac{g}{2\sqrt{2}}W^-_\mu \sum_{\ell=e,\mu,\tau}
\bar \nu_\ell\gamma^\mu\left(1-\gamma_5\right)\ell
\ea
C and P are broken by the $(1-\gamma_5)$ in the couplings.
CP is broken if $V_{CKM}=\left(V_{ij}\right)$ is (irreducibly) {\em
complex}.

\begin{floatingfigure}[r]{8cm}
\setlength{\unitlength}{3pt}
\begin{picture}(60,40)(0,-30)
\SetScale{3}
\SetWidth{0.75}
\ArrowLine(0,0)(20,0)\Text(0,5)[lb]{ $\mu^-$}
\Vertex(20,0){1.5}
\ArrowLine(20,0)(40,10)\Text(40,10)[l]{ $\nu_\mu$}
{\Photon(17,0)(33,-20){1.5}{4}
\Text(26,-5)[lb]{$W^-$}}
\Vertex(30,-20){1.5}
\ArrowLine(30,-20)(50,-10)\Text(50,-10)[l]{ $e^-$}
\ArrowLine(50,-30)(30,-20)\Text(50,-30)[l]{ $\bar\nu_e$}
\end{picture}
\caption{Muon decay: the main source of our knowledge of $g$.}
\end{floatingfigure}
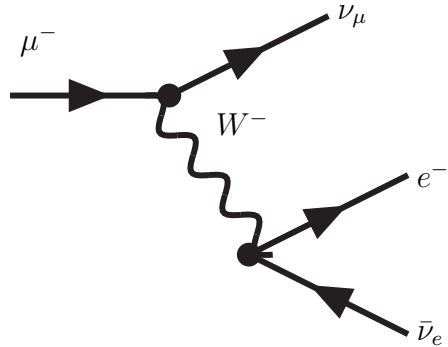
The couplings constant $g$ together with the mass $M_W^2$ can be 
determined from the Fermi constant as measured in muon decay.
The relevant corrections are known to two-loop level. The most recent
calculations and earlier references can be found in \cite{muon}.
The result is
\be
 G_F = \frac{g^2}{8M_W^2} = 1.16639(1)\cdot10^{-5}~\mbox{GeV}^{-2}\,.
\ee

The Cabibbo-Kobayashi-Maskawa matrix
$V_{CKM}=\left(V_{ij}\right)$ results from diagonalizing the quark mass
terms resulting from the Yukawa terms and the Higgs vacuum expectation value.
It is a general unitary matrix but the phases of the quark fields
can be redefined. This allows to remove 5 of the 6 phases present
in a general unitary 3 by 3 matrix\footnote{We have 6 quark fields
but changing all the anti-quarks by a phase and the quarks by the opposite
phase results in no change in $V_{CKM}$.}.
The matrix $V_{CKM}$ thus contains three phases and one mixing angle.
The Particle Data Group preferred parametrization is \cite{PDG}
\be
 \left(\begin{array}{ccc}
c_{12}c_{13} & s_{12}c_{13} & s_{13}e^{-i\delta_{13}}\\
-s_{12}c_{23}-c_{12}s_{23}s_{13}e^{i\delta_{13}} &
c_{12}c_{23}-s_{12}s_{23}s_{13}e^{i\delta_{13}} & s_{23}c_{13}\\
s_{12}s_{23}-c_{12}c_{23}s_{13}e^{i\delta_{13}} &
-c_{12}s_{23}-s_{12}c_{23}s_{13}e^{i\delta_{13}} & c_{23}c_{13}\,.
\end{array}\right)
\ee
Using the measured experimental values, see below and \cite{Grinstein},
\be
s_{12}=\sin\theta_{12}\approx0.2;
s_{23}\approx0.04\mbox{ and }s_{13}\approx0.003
\ee
an approximate parametrization, known as the
Wolfenstein parametrization, can be given.
This is defined via
$s_{12}c_{13}\equiv\lambda$;
$
s_{23}c_{13}\equiv A\lambda^2$
and
$s_{13}e^{-i\delta_{13}}\equiv A\lambda^3(\rho-i\eta)$.
To order $\lambda^4$ the CKM-matrix is
\be
\left(\begin{array}{ccc}
1-\frac{\lambda^2}{2} & \lambda & A\lambda^3(\rho-i\eta)\\
-\lambda & 1-\frac{\lambda^2}{2} & A\lambda^2\\
A\lambda^3(1-\rho-i\eta) & -A\lambda^2 & 1
\end{array}\right)
\ee

\section{Semi-leptonic Decays}
\subsection{$V_{ud}$}

The 
underlying idea always the same.
The vector current has matrix-element one at zero momentum transfer
because of the underlying vector Ward identities.
The element $V_{ud}$ is measured in three main sets of decays.
\begin{itemize}
\item{\bf neutron:}
Here we use in addition 
that the axial current effect is actually measurable via
the angular distribution of the electron. So
$\Gamma(n\to pe^-\bar\nu) \sim G_n^2 (1+3g_A^2/g_V^2) A$\\
with $A$ calculable but containing photon loops.
Using the value for $g_A/g_V$ and the neutron lifetime
of the 98 particle data book\cite{PDG98}
and the calculations of radiative corrections quoted in \cite{woolcock}
I obtain
\be
|V_{ud}| = 0.9792(40)\,.
\ee
At present the errors are dominated by the measurement of $g_A/g_V$.
This could change in the near future.

\item{\bf Nuclear superallowed $\beta$-decays: $0^+\to0^+$}
The main advantages here are that only the vector current can contribute
and that very accurate experimental results are available.
The disadvantage is that the charge symmetry breaking, or isospin, effects
and the photonic radiative corrections are nuclear structure dependent
with an unknown error. The quoted theory errors are such that
the measurements for different nuclei are in contradiction
with each other. In \cite{PDG98} they therefore quote
\be
|V_{ud}| = 0.9740(10)\,.
\ee
\item{\bf Pion $\beta$-decay: $\pi^+\to\pi^0 e^+\bar\nu$}
The theory here is very clean and improvable using CHPT.
The disadvantage is that the 
branching ratio is about $10^{-8}$, known to
about 4\%. Experiments at PSI are in progress to get 0.5\%.
\end{itemize}
In the future better experimental precision
in neutron and $\pi$ $\beta$-decay should improve the determination.
$|V_{ud}|$ is by far the best directly determined CKM-matrix element.

\subsection{$V_{us}$}

Again we use the fact that the matrix-element
of a conserved vector current is one at zero momentum. But compared to the
previous subsection we have additional complications:\\[0.2mm]
$\bullet$ The vector current is $\bar s\gamma_\mu u$ so corrections
are $(m_s-m_u)^2$ and not $(m_d-m_u)^2$ so they are naively larger.\\
$\bullet$ A longer extrapolation to the zero-momentum point is
needed.\\[0.2mm]
The relevant semi-leptonic decays can be measured in hyperon or in Kaon
decays.

{\bf Hyperon $\beta$-decays (e.g. $\Sigma^-\to n e^-\bar\nu,
\Lambda\to p e^-\bar\nu$)}:
 Here there are large theoretical problems.
In CHPT the corrections are large and many new parameters show up. In
addition the series does not converge too well. Curiously enough,
using lowest order CHPT with model-corrections works OK.
For references consult the relevant section of \cite{PDG,PDG98}.
This area needs theoretical work very badly.

{\bf Kaon $\beta$-decays (e.g. $K^+\to \pi^0 e^+\nu$)}:
Both theory and data are old by now. The theory was done by
Leutwyler and Roos \cite{LW}. The analysis uses old-fashioned
photonic loops for the electro-magnetic corrections and one-loop CHPT
for the strong corrections due to quark masses.
Both aspects are at present being improved \cite{kl3improve}.
The latest data are from 1987 and the most precise ones are older.
There is at present a proposal at BNL while KLOE at DAPHNE should also
be able to improve the precision.

The result obtained from the last process is
\be
|V_{us}| = 0.2196\pm0.0023\,.
\ee
We can now use this to test the unitarity relation
\be
|V_{us}|^2+|V_{ud}|^2 = 0.9969\pm0.0022
\ee
$|V_{ub}|^2$ is so small it is not visible in the precision shown here.
This is a small discrepancy which should be understood but no real cause
for worry at present.

\subsection{Testing CHPT/QCD and determining CHPT parameters}

In the sector of semi-leptonic Kaon decays CHPT unfolds all of its
power. An extensive review can be found in \cite{semidaphne}
with a more recent update in \cite{kaon99}. There are of course also
numerous model calculations and other approaches existing.
An example of model calculations using Schwinger-Dyson
equations is in \cite{semimodel}. Notice that the Schwinger-Dyson equations
themselves do not constitute the model aspect but the assumptions
made in their solutions.

I now simply list the main decays and which quantity they test and/or measure.
\begin{itemize}
\item$K\to\mu\nu$ ($K_{\ell2}$) : Measurement of $F_K$.
This known to two loops\cite{ABT1} in CHPT and the electromagnetic corrections
have also been updated \cite{Knecht1}

\item
$K\to\pi\ell\nu$ ($K_{\ell3}$): 
$V_{us}$ and form-factors, see \cite{semidaphne} and references
therein.
\item
$K\to\pi\pi\ell\nu$ ($K_{\ell4}$): Form-factors and a main source of
CHPT input parameters. Known at two-loops in CHPT \cite{ABT2}.
\item
$K\to\pi\ell\nu\gamma$ ($K_{\ell3\gamma}$): Lots of form-factors with
large corrections combining to a final small correction\cite{BEG}.
\item
$K(\pi)\to\ell\nu\gamma$ ($K(\pi)_{\ell2\gamma}$):
This has two form-factors, one normal and one anomalous. The interference
between them allows to see the sign of the anomaly. This can also be done
in $K_{\ell4}$ and both confirm nicely the
expectations\cite{BEG,semidaphne,ABBC}.
\end{itemize}

\section{Non-leptonic Decays: $K\to\pi\pi$ and $K$-$\kob$ mixing}

We have been rather successful in understanding the theory behind the
semi-leptonic decays discussed in the previous section. Basically we always
used CHPT or similar arguments to get at the coupling of the $W$-boson
to hadrons. A similar simple approach fails completely for non-leptonic
decays. I'll first discuss the main qualitative problem that shows up
in trying to estimate these decays, then the phenomenology involved
in mixing phenomena and then proceed in the various steps needed
to actually calculate these processes in the standard model.
Finally some numerical results are presented.

\subsection{The $\Delta I=1/2$ rule}

\begin{figure}[t]%{10cm}
%\raisebox{-65pt}{
\phantom{p}\hskip2cm
\SetScale{0.7}
\setlength{\unitlength}{0.7pt}
\begin{picture}(200,200)(0,0)
\SetWidth{2}
\Text(0,105)[lb]{$K^+$}
\Line(0,100)(60,100)
\Text(110,155)[rb]{$\pi^0$}
\Line(60,100)(110,150)
\Photon(60,100)(140,50){10}{3}
\Text(120,80)[lb]{$W^+$}
\Vertex(60,100){7}
\Vertex(140,50){7}
\Text(200,45)[rt]{$\pi^+$}
\Line(140,50)(200,50)
\end{picture}
%}%raisebox
%\raisebox{-65pt}{
\hskip1cm
\SetScale{0.7}
\setlength{\unitlength}{0.7pt}
\begin{picture}(200,200)(0,0)
\SetWidth{2}
\Text(0,105)[lb]{$K^+$}
\Line(0,100)(60,100)
\Photon(60,100)(140,100){10}{3}
\Text(100,120)[lb]{$W^+$}
\Vertex(60,100){7}
\Vertex(140,100){7}
\Text(200,45)[rt]{$\pi^0$}
\Line(140,100)(200,50)
\Text(200,155)[rb]{$\pi^+$}
\Line(140,100)(200,150)
\end{picture}
%}%raisebox
\caption{The two naive $W^+$-exchange diagrams for
$K^+\longrightarrow \pi^+\pi^0$.}
\label{figkpipiplus}
\end{figure}
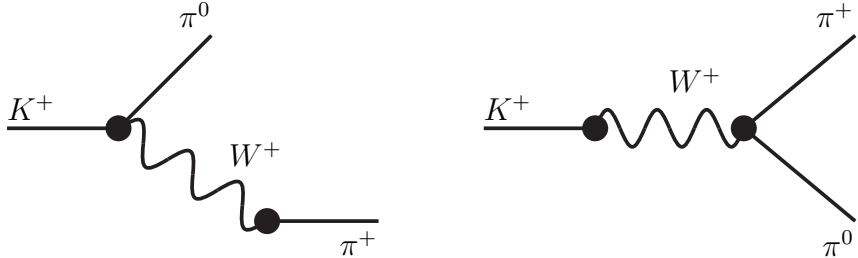

The underlying problem appears when we try to calculate $K\to\pi\pi$
decays in a similar fashion as for the semi-leptonic decays.
For $K^+\to\pi^+\pi^0$ we can draw two Feynman diagrams with a simple $W^+$
exchange as shown in Fig. \ref{figkpipiplus}.
The relevant $W^+$-hadron couplings have all been measured in semi-leptonic
decays and so we have a unique prediction. Comparing this with the
measured decay we get within a factor of two or so.

A much worse result appears when we try the same method for the neutral decay
$K^0\to\pi^0\pi^0$. As shown in Fig. \ref{figkpipizero} there is no
possibility to draw diagrams similar to those in Fig. \ref{figkpipiplus}.
The needed vertices always violate charge-conservation.
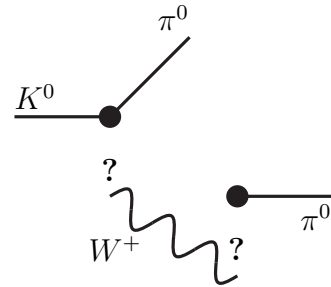
\begin{floatingfigure}[r]{5cm}
%\raisebox{-65pt}
{
\SetScale{0.6}
\setlength{\unitlength}{0.6pt}
\begin{picture}(200,160)(0,0)
\SetWidth{2}
\Text(0,105)[lb]{$K^0$}
\Line(0,100)(60,100)
\Vertex(60,100){7}
\Text(110,155)[rb]{$\pi^0$}
\Line(60,100)(110,150)
\Photon(60,50)(140,0){10}{3}
\Text(80,25)[rt]{$W^+$}
\Text(60,60)[b]{\bf ?}
\Text(140,10)[b]{\bf ?}
\Vertex(140,50){7}
\Text(200,45)[rt]{$\pi^0$}
\Line(140,50)(200,50)
\end{picture}
}%raisebox%
\caption{No simple $W^+$-exchange diagram is possible for
$K^0\longrightarrow \pi^0\pi^0$.}
\label{figkpipizero}
\end{floatingfigure}
So we expect that the neutral decay should be small compared
with the ones with charged pions. Well, if we look at the experimental
results we see
\ba
\Gamma(K^0\longrightarrow\pi^0\pi^0)=&
\dsp\frac{1}{2}\Gamma(K_S\longrightarrow\pi^0\pi^0)&=2.3\;10^{-12}\mbox{ MeV}
\nonumber\\
\Gamma(K^+\longrightarrow\pi^+\pi^0)=&\dsp
=1.1\;10^{-14}\mbox{ MeV}&
\ea
So the expected zero one is by far the largest !!!

The same conundrum can be expressed in terms of the
 isospin amplitudes:\footnote{Here there are several different sign
and normalization conventions
possible. I present the one used in the work by J.~Prades and myself.}
\ba
A[K^0 \to \pi^0 \pi^0]  \equiv \sqrt{\frac{1}{3}} { A_0}
-\sqrt{\frac{2}{3}} \, { A_2}&& \nonumber\\ 
A[K^0 \to \pi^+ \pi^-]  \equiv \sqrt{\frac{1}{3}} { A_0}
+\frac{1}{\sqrt 6} \, { A_2} &&
A[K^+ \to \pi^+ \pi^0]  \equiv \frac{\sqrt 3}{2} { A_2}\,.
\ea
The above quoted experimental results can now be rewritten as
\be 
\left|\frac{A_0}{A_2}\right| = 22.1
\ee
while the naive $W^+$-exchange discussed would give
$
\left|{A_0}/{A_2}\right| = \sqrt{2}\,.
$
This discrepancy is known as the problem of the $\Delta I=1/2$ rule.
The amplitude which changes the isospin $1/2$ to zero is much larger than
the one that changes the isospin to 2 by $3/2$.

Some enhancement is easy to understand from final state $\pi\pi$-rescattering.
Removing these and higher order effects in the light quark masses
one obtains\cite{KMW}
\be
\dsp\left|\frac{A_0}{A_2}\right| = 16.4\,.
\ee

Later we also need the amplitudes with the final state interaction
phase removed via
\be
A_I=-ia_I e^{i\delta_I}
\ee
for $I=0,2$. $\delta_I$ is the angular momentum zero isospin I scattering
phase at the Kaon mass.

\subsection{Phenomenology of $K$-$\kob$ mixing}

The $K^0$ and \kob\ states are the ones with $\bar sd$ and $\bar ds$
quark content respectively. Up to free phases in these states we can
define the action of $CP$ on these states as
\be
CP |K^0\rangle = -|\kob\rangle\,.
\ee
{}We can construct eigenstates with a definite $CP$ transformation:
\be
K^0_{1(2)} = \frac{1}{\sqrt{2}}
\left(K^0 -(+) \kob\right)\quad\quad CP|K_{1(2)} = +(-)|K_{1(2)}\,.
\ee
Now the main decay mode of $K^0$-like states is $\pi\pi$. A two pion state
with charge zero in spin zero is always CP even.
Therefore the decay $K_1\to\pi\pi$ is possible
but $K_2\to\pi\pi$ is {\em impossible}; $K_2\to\pi\pi\pi$ is possible.
Phase-space for the $\pi\pi$ decay is much larger than for the
three-pion final state. Therefore if we start out with a pure $K^0$ or \kob\
state its
$\Rightarrow$ $K_2$ lives much longer than $K_1$.
So after a, by microscopic standards, long time only the $K_2$ component
survives.

In the early sixties, it pays off to do precise experiments,
one actually measured \cite{CCFT}
\be
\frac{\Gamma(K_L\to\pi^+\pi^-)}{\Gamma(K_L\to\mbox{all})}=
(2\pm0.4)\cdot10^{-3}\,.
\ee
So we see that \fbox{\large CP is violated}\,.

This leaves us with the questions:
\begin{itemize}
\item Does $K_1$ turn in to $K_2$ (mixing or indirect CP violation) ?
\item Does $K_2$ decay directly into $\pi\pi$ (direct CP violation) ?
\end{itemize}
In fact, the 
answer to both is {\em YES} and is major qualitative test
of the standard model Higgs-fermion sector.

Let us now describe the $K^0\kob$ system in somewhat more detail.
The Hamiltonian, seen  as a two state system, is given by
\be
i\frac{d}{dt}
\left(\begin{array}{c}K^0\\\overline K^0\end{array}\right)
=
\left(\begin{array}{cc}M_{11}-\frac{i}{2}\Gamma_{11}&
M_{12}-\frac{i}{2}\Gamma_{12}\\
M_{21}-\frac{i}{2}\Gamma_{21}&M_{22}-\frac{i}{2}\Gamma_{22}\end{array}\right)
\left(\begin{array}{c}K^0\\\overline K^0\end{array}\right)
\ee
where $M = \left(M_{ij}\right)$ and $\Gamma =\left(\Gamma_{ij}\right)$
are hermitian two by two matrices. The Hamiltonian itself is allowed to have a
non-hermitian part since we do not conserve probability here. The Kaons
themselves can decay and the anti-hermitian part $\Gamma$ describes the
decays of the Kaons to the ``rest of the universe.''

CPT implies \cite{eduardo}
\be
M_{11} = M_{22}\quad \Gamma_{11}=\Gamma_{22}\quad M_{12}=M_{21}^*\quad
\Gamma_{12}=\Gamma_{21}^*
\ee
and this assumption can in fact be relaxed for tests of CPT.

Diagonalizing the Hamiltonian we obtain
\be
K_{S(L)} = \frac{1}{\sqrt{1+|\tilde\varepsilon|^2}}
\left(K_{1(2)}+\tilde\varepsilon K_{2(1)}\right)
\ee
as physical propagating states. Notice that they are not orthogonal.

For observables we now define:
\be
\varepsilon = \frac{A(K_L\to(\pi\pi)_{I=0})}{A(K_S\to(\pi\pi)_{I=0})}
\,,\quad
\eta_{+-} = \frac{A(K_L\to\pi^+\pi^-)}{A(K_S\to\pi^+\pi^-)}
\,,\quad
\,,\quad
\eta_{00} = \frac{A(K_L\to\pi^0\pi^0)}{A(K_S\to\pi^0\pi^0)}
\ee
and
\be
\varepsilon^\prime = \frac{1}{\sqrt{2}}\left(
\frac{A(K_L\to(\pi\pi)_{I=2})}{A(K_S\to(\pi\pi)_{I=0})}-
\varepsilon
\frac{A(K_S\to(\pi\pi)_{I=2})}{A(K_S\to(\pi\pi)_{I=0})}\right)\,.
\ee
The latter has been specifically constructed to remove the 
$K^0$-\kob\ transition.
$|\varepsilon|$, $|\eta_{+-}|$ and
$|\eta_{00}|$ are directly measurable.

We can now make a series of approximations that are experimentally
valid,
\be
|\re a_0| >> |\re a_2| >> |\im a_0|,|\im a_2| \quad\quad\quad
 |\varepsilon|,|\tilde \varepsilon| << 1 \quad |\varepsilon^\prime| <<
| \varepsilon|\,,
\ee
to obtain the usually quoted expressions
\be
\varepsilon^\prime = \frac{i}{\sqrt{2}}e^{i(\delta_2-\delta_0)}
\frac{\re a_2}{\re a_0}\left(\frac{\im a_2}{\re a_2}-\frac{\im a_0}{\re a_0}
\right)
\quad\mbox{and}\quad
\varepsilon = \tilde\varepsilon+ i\frac{\im a_0}{\re a_0}\,.
\ee
For the latter we
use $\Delta m = m_L-m_S \approx \frac{\Delta\Gamma}{2}$ and 
$\Gamma_L << \Gamma_S$ and the fact that
$\Gamma_{12}$ is dominated by $\pi\pi$ states and get
\be
\varepsilon = \frac{1}{\sqrt{2}}e^{i\pi/4}
\left(\frac{\im M_{12}}{\Delta m}+\frac{\im a_0}{\re a_0}\right)\,.
\ee
Putting all the above together we finally get to
\be
\eta_{+-} = \varepsilon+\varepsilon^\prime
\quad\mbox{and}\quad
\eta_{00} = \varepsilon-2\varepsilon^\prime\nonumber
\ee
Where you can see that $\varepsilon$ describes the indirect part
and $\varepsilon^\prime$ the direct part, since the mixing contribution would
be the same for $\eta_{+-}$ and $\eta_{00}$.

\subsection{Experimental results and SM-diagrams}

\begin{table}
\hskip5cm
{\begin{tabular}{|cc|}
\hline
NA31 & $(23.0\pm6.5)\times 10^{-4}$\\
E731 & $(7.4\pm5.9)\times 10^{-4}$\\
\hline
KTeV & $(28.0\pm4.1)\times10^{-4}$\\
NA48 97 & $(18.5\pm7.3)\times10^{-4}$\\
NA48 98 & $(12.2\pm4.9)\times10^{-4}$\\
\hline
ALL & $(19.3\pm2.4)\times10^{-4}$\\
\hline
\end{tabular}}
\caption{Recent results on $\varepsilon^\prime/\varepsilon$. The total
$\chi^2$ of the fit is $\chi^2/dof = 11.1/5$.}
\label{tabepspeps}
\end{table}
Recent experimental results are
\be
|\varepsilon| = 2.28\cdot 10^{-3}
\quad\mbox{and for}\quad
\re\left(\frac{\varepsilon^\prime}{\varepsilon}\right)
= \frac{1}{6}\left\{
1-\left|\frac{\eta_{00}}{\eta_{+-}}\right|^2\right\}
\ee
we show the recent results in Table \ref{tabepspeps}.
The data are taken from \cite{epspeps}.

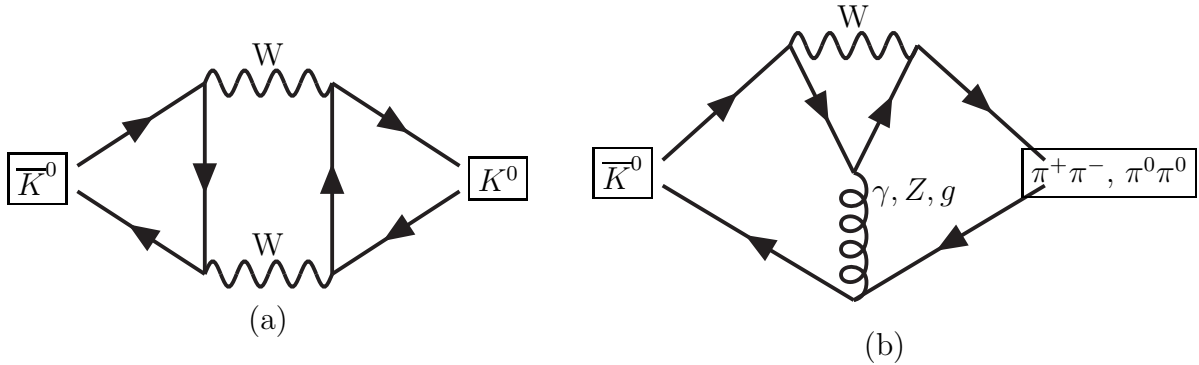
\begin{figure}
\setlength{\unitlength}{2.4pt}
\SetScale{2.4}
\SetWidth{0.6}
\begin{picture}(100,50)(-10,-25)
\Text(0,0)[r]{\fbox{$\overline K^0$} }
\ArrowLine(0,2)(20,15)
\ArrowLine(20,-15)(0,-2)
\ArrowLine(20,15)(20,-15)
\Photon(20,15)(40,15){2}{4}\Text(30,18)[b]{{W}}
\Photon(20,-15)(40,-15){2}{4}\Text(30,-12)[b]{{W}}
\ArrowLine(40,15)(60,2)
\ArrowLine(40,-15)(40,15)
\ArrowLine(60,-2)(40,-15)
\Text(60,0)[l]{ \fbox{$K^0$}}
\Text(30,-25)[b]{(a)}
\end{picture}
\hskip-0.5cm
\raisebox{0.5cm}{
\setlength{\unitlength}{2.4pt}
\SetScale{2.4}
\SetWidth{0.6}
\begin{picture}(70,50)(0,-20)
\Text(0,0)[r]{\fbox{$\overline K^0$} }
\ArrowLine(0,2)(20,20)
\ArrowLine(30,-20)(0,-2)
{
\Photon(20,20)(40,20){2}{4}}\Text(30,23)[b]{{W}}
\ArrowLine(20,20)(30,0)
\ArrowLine(30,0)(40,20)
\ArrowLine(40,20)(60,2)
\ArrowLine(60,-2)(30,-20)
\Text(55,0)[l]{ \fbox{$\pi^+\pi^-$, $\pi^0\pi^0$}}
{
\Gluon(30,0)(30,-20){2}{4}}\Text(33,-3)[l]{{$\gamma,Z,g$}}
\Text(35,-30)[b]{(b)}
\end{picture}
}%raisebox
\caption{(a) The box diagram contribution to $K^0$\kob\ mixing.
Extra gluons etc. are not shown.
(b) The Penguin.diagram contribution to $K\to\pi\pi$.}
\label{figbox}
\end{figure}

In the standard model $K^0$\kob\ mixing comes from the box-diagram
of Fig. \ref{figbox}(a)
Since it contains $W$-couplings to all generations CP-violation from
the CKM matrix is possible in this contribution.
The Penguin diagram shown in Fig. \ref{figbox}(b) contributes to the direct
CP-violation as given by $\varepsilon^\prime$.
Again, $W$-couplings to all three generations show up so CP-violation
is possible in $K\to\pi\pi$. This is a qualitative prediction of the
standard model and borne out by experiment.

\subsection{The steps from quarks to mesons in the weak interaction}

The full calculation can be described by three steps depicted in
Fig. \ref{figsteps}.
\begin{figure}
\hskip1.5cm
\begin{tabular}{ccc}
ENERGY SCALE & FIELDS & Effective Theory\\
\hline
\\[2mm]
$M_W$ & \framebox{\parbox{5cm}{\begin{center}
$W,Z,\gamma,g$;\\$\tau,\mu,e,\nu_\ell$;\\
$t,b,c,s,u,d$\end{center}
}} & Standard Model\\[2mm]
 & {\em $\Downarrow$ using OPE} & \\[2mm]
$\lesssim m_c$ & \framebox{\parbox{5cm}{\begin{center}
$\gamma,g$; $\mu,e,\nu_\ell$;\\ $s,d,u$\end{center}
}} & QCD,QED,${\cal H}_{\mbox{eff}}^{|\Delta S|=1,2}$\\[2mm]
  & {\em$\Downarrow$ ???} & \\[2mm]
$M_K$  & \framebox{\parbox{5cm}{\begin{center}$\gamma$; $\mu,e,\nu_\ell$;\\
$\pi$, $K$, $\eta$\end{center}}} & CHPT\\ \\\hline
\end{tabular}
\caption{A schematic exposition of the various steps in the calculation
of nonleptonic matrix-elements.}
\label{figsteps}
\end{figure}
We cannot simply use one-loop perturbation theory for the short-distance
part since
$\log{\frac{M_W^2}{M_K^2}}$ and
$\log{\frac{M_t^2}{M_K^2}}$ are large and their
effects need to be resummed. This can be done using
OPE and renormalization group methods.

First we integrated out the heaviest particles step by step using Operator
Product Expansion methods. The steps OPE we descibe in the next subsections
while step {\em ???} we will split up in more subparts later.

\subsection{Step I: from SM to OPE}

The contribution from the standard model diagrams in Fig. \ref{figsSM}
\begin{figure}
\begin{minipage}{0.50\textwidth}
\vskip-4cm
\setlength{\unitlength}{1.8pt}
\begin{picture}(60,50)(0,25)
\SetScale{1.8}
\SetWidth{0.75}
\ArrowLine(10,0)(0,20)\Text(0,20)[r]{$s$ }
\ArrowLine(0,-20)(10,0)\Text(0,-20)[r]{$u,c$ }
\Vertex(10,0){2}
{\Photon(10,0)(30,0){1}{3}\Text(20,3)[b]{{$W$}}}
\Vertex(30,0){2}
\ArrowLine(40,20)(30,0)\Text(40,20)[l]{ $d$}
\ArrowLine(30,0)(40,-20)\Text(40,-20)[l]{ $u,c$}
\end{picture}
\setlength{\unitlength}{1.8pt}
\begin{picture}(60,50)(0,25)
\SetScale{1.8}
\SetWidth{0.75}
\ArrowLine(10,0)(0,20)\Text(0,20)[r]{$s$ }
\ArrowLine(0,-20)(10,0)\Text(0,-20)[r]{$u,c$ }
\Vertex(10,0){2}
\Vertex(2.5,-15){2}
\Vertex(37.5,-15){2}
{\Photon(10,0)(30,0){1}{3}\Text(20,3)[b]{{$W$}}}
\Vertex(30,0){2}
\ArrowLine(40,20)(30,0)\Text(40,20)[l]{ $d$}
\ArrowLine(30,0)(40,-20)\Text(40,-20)[l]{ $u,c$}
{
\Gluon(2.5,-15)(37.5,-15){-2}{5}
\Text(20,-19)[t]{{$\gamma,g,Z$}}}
\end{picture}
\\
\vskip-1cm
\setlength{\unitlength}{1.8pt}
\begin{picture}(60,50)(0,25)
\SetScale{1.8}
\SetWidth{0.75}
\ArrowLine(10,0)(0,10)\Text(0,10)[r]{$s$ }
\ArrowLine(20,-20)(10,0)
\ArrowLine(30,0)(20,-20)\Text(27,-10)[l]{ $u,c$}
\Vertex(10,0){2}
\Vertex(30,0){2}
\Vertex(20,-20){2}
\ArrowLine(40,10)(30,0)\Text(40,10)[l]{ $d$}
\ArrowLine(0,-40)(20,-40)
\ArrowLine(20,-40)(40,-40)
\Text(40,-40)[l]{ $u,c$}
\Vertex(20,-40){2}
{
\Photon(10,0)(30,0){1}{3}\Text(20,3)[b]{{$W$}}
}
{
\Gluon(20,-20)(20,-40){2.5}{3}
\Text(22,-30)[l]{ {$\gamma,g,Z$}}
}
\end{picture}
%\hfill
\setlength{\unitlength}{1.8pt}
\begin{picture}(60,50)(0,25)
\SetScale{1.8}
\SetWidth{0.75}
\ArrowLine(10,0)(0,10)\Text(0,10)[r]{$s$ }
\Vertex(10,0){2}
\Vertex(30,0){2}
\Vertex(20,-20){2}
\ArrowLine(40,10)(30,0)\Text(40,10)[l]{ $d$}
\ArrowLine(0,-40)(20,-40)
\ArrowLine(20,-40)(40,-40)
\Text(40,-40)[l]{ $u,c$}
{
\ArrowLine(20,-20)(10,0)
\ArrowLine(30,0)(20,-20)\Text(27,-10)[l]{ {$t$}}}
{
\Photon(10,0)(30,0){1}{3}\Text(20,3)[b]{{$W$}}}
{
\Gluon(20,-20)(20,-40){2.5}{3}
\Text(22,-30)[l]{ {$\gamma,g,Z$}}}
\Vertex(20,-40){2}
\end{picture}
\vskip5cm
\caption{The standard model diagrams to be calculated at a high scale.}
\label{figsSM}
\end{minipage}
\hspace{0.02\textwidth}
\begin{minipage}{0.47\textwidth}
\setlength{\unitlength}{1.8pt}
\begin{picture}(40,40)(0,-20)
\SetScale{1.8}
\SetWidth{0.75}
\ArrowLine(20,0)(0,20)\Text(0,20)[r]{$s$ }
\ArrowLine(40,20)(20,0)\Text(40,20)[l]{ $d$}
\ArrowLine(0,-20)(20,0)
\ArrowLine(20,0)(40,-20)
\Text(20,0)[]{$\bigotimes$}
\end{picture}
%\hskip0.5cm
\raisebox{-1cm}{
\setlength{\unitlength}{1.6pt}
\begin{picture}(40,40)(0,-40)
\SetScale{1.6}
\SetWidth{1.1}
\ArrowLine(20,0)(0,20)\Text(0,20)[r]{$s$ }
\ArrowLine(40,20)(20,0)\Text(40,20)[l]{ $d$}
\ArrowLine(0,-40)(20,-40)
\ArrowLine(20,-40)(40,-40)
\Text(20,0)[]{$\bigotimes$}
\ArrowArc(20,-10)(10,90,270)
\ArrowArc(20,-10)(10,-90,90)
\Vertex(20,-20){2}
{
\Gluon(20,-20)(20,-40){2.5}{3}
\Text(22,-30)[l]{ {$\gamma,g$}}}
\Vertex(20,-40){2}
\end{picture} }
\setlength{\unitlength}{1.8pt}
\begin{picture}(40,40)(0,-20)
\SetScale{1.8}
\SetWidth{0.75}
\ArrowLine(20,0)(0,20)\Text(0,20)[r]{$s$ }
\ArrowLine(40,20)(20,0)\Text(40,20)[l]{ $d$}
\ArrowLine(0,-20)(20,0)
\ArrowLine(20,0)(40,-20)
\Text(20,0)[]{$\bigotimes$}
\Vertex(5,-15){2}
\Vertex(35,-15){2}
{\Gluon(5,-15)(35,-15){-2}{5}}
\Text(20,-19)[t]{{$\gamma,g$}}
\end{picture}
%\hskip0.5cm
\caption{The diagrams needed for the
matrix-elements calculated at a scale $\mu_H\approx m_W$
using the effective Hamiltonian.}
\label{figsOPE}
\end{minipage}
\end{figure}
we now replace with a contribution of an effective Hamiltonian given by
\be
 {\cal H}_{\mbox{eff}} = \sum_i C_i(\mu) Q_i(\mu)
\label{defHeff}
\dsp{\cal H}_{\mbox{eff}} = \frac{G_F}{\sqrt{2}} V_{ud}V_{us}^*
\sum_i\left(z_i-y_i\frac{V_{td}V_{ts}^*}{V_{ud}V_{us}^*}\right)Q_i\,.
\ee
In the last part we have real coefficients $z_i$ and $y_i$ and
 the CKM-matrix elements occurring are shown explicitly.
The four-quark operators $Q_i$ are defined by
\ba
\label{defQi}
Q_1^u = 
(\bar s_\alpha\gamma_\mu u_\beta)_L (\bar u_\beta\gamma^\mu d_\alpha)_L
&&
Q_1^c = (\bar s_\alpha\gamma_\mu c_\beta)_L (\bar c_\beta\gamma^\mu d_\alpha)_L
\nonumber\\
Q_2^u =
 (\bar s_\alpha\gamma_\mu u_\alpha)_L (\bar u_\beta\gamma^\mu d_\beta)_L
&&
Q_2^c = (\bar s_\alpha\gamma_\mu c_\alpha)_L (\bar c_\beta\gamma^\mu d_\beta)_L
\nonumber\\
Q_3 = (\bar s_\alpha\gamma_\mu d_\alpha)_L
\sum_{q=u,d,s,c,b} (\bar q_\beta\gamma^\mu q_\beta)_L
&&
Q_4 = (\bar s_\alpha\gamma_\mu d_\beta)_L
\sum_{q=u,d,s,c,b} (\bar q_\beta\gamma^\mu q_\alpha)_L
\nonumber\\
Q_5 = (\bar s_\alpha\gamma_\mu d_\alpha)_L
\sum_{q=u,d,s,c,b} (\bar q_\beta\gamma^\mu q_\beta)_R
&&
Q_6 = (\bar s_\alpha\gamma_\mu d_\beta)_L
\sum_{q=u,d,s,c,b} (\bar q_\beta\gamma^\mu q_\alpha)_R
\nonumber\\
Q_7 = (\bar s_\alpha\gamma_\mu d_\alpha)_L
\sum_{q=u,d,s,c,b} \frac{3}{2}e_q(\bar q_\beta\gamma^\mu q_\beta)_R
&&
Q_8 = (\bar s_\alpha\gamma_\mu d_\beta)_L
\sum_{q=u,d,s,c,b} \frac{3}{2}e_q(\bar q_\beta\gamma^\mu q_\alpha)_R
\nonumber\\
Q_9 = (\bar s_\alpha\gamma_\mu d_\alpha)_L
\sum_{q=u,d,s,c,b} \frac{3}{2}e_q(\bar q_\beta\gamma^\mu q_\beta)_L
&&
Q_{10} = (\bar s_\alpha\gamma_\mu d_\beta)_L
\sum_{q=u,d,s,c,b} \frac{3}{2}e_q(\bar q_\beta\gamma^\mu q_\alpha)_L
\ea
with $(\bar q\gamma_\mu q^\prime)_{(L,R)} 
= \bar q \gamma_\mu(1\mp\gamma_5)q^\prime$;
$\alpha$ and $\beta$ are colour indices.

So we calculate now matrix elements between quarks and gluons in the
standard model using the diagrams of Fig. \ref{figsSM}
and equate those to the same matrix-elements calculated using the
effective Hamiltonian of Eq. \ref{defHeff} and the diagrams
of Fig. \ref{figsOPE}.

{\underline{Notes}:}\\
$\bullet$ In the Penguins CP-violation shows up since all 3 generations
are present.\\
$\bullet$ The equivalence done by matrix-elements between
{\em Quarks and Gluons}\\
$\bullet$ The SM part is $\mu_W$-independent to $\alpha_S^2(\mu_W)$.\\
$\bullet$ OPE part: The $\mu_H$ dependence of
$C_i(\mu_H)$ cancels the $\mu_H$ dependence of the diagrams
to order $\alpha_S^2(\mu_H)$.

This gives at $\mu_H=M_W$ in the NDR-scheme\footnote{The precise definition
of the four-quark operators $Q_i$ comes in here as well. See the lectures by
Buras \cite{Buras1} for a more extensive description of that.}
the numerical values give in Table \ref{tabzimw}.
\begin{floatingtable}{
\begin{tabular}{|ccc|}
\hline
$z_1$   & 0.053     &  { $g,\gamma$-box} \\
$z_2$	& 0.981	    &{ $W^+$-exchange  $g,\gamma$-box} \\
\hline
$y_3$	& 0.0014    & {$g,Z$-Penguin  $WW$-box} \\
$y_4$	&$-$0.0019  & {$g$-Penguin}\\
$y_5$	& 0.0006    & {$g$-Penguin}\\
$y_6$	&$-$0.0019  & {$g$-Penguin}\\
$y_7$	& 0.0009    & {$\gamma,Z$-Penguin}\\
$y_8$	& 0.	    & \\
$y_9$	& $-$0.0074 & {$\gamma,Z$-Penguin  $WW$-box}\\
$y_{10}$& 0.        & \\
\hline
\end{tabular}}
\caption{The Wilson coefficients and their main source at
the scale $\mu_H=m_W$ in the NDR-scheme.}
\label{tabzimw}
\end{floatingtable}
In the same table I have
given the main source of these numbers. Pure tree-level $W$-exchange
would have only given $z_2=1$ and all others zero.
Note that the coefficients from $\gamma,Z$ exchange are similar to the gluon
exchange ones.

Now comes the main advantage of the OPE formalism. Using the
renormalization group equations we can now calculate
the change with $\mu$ of the $C_i$ thus resumming the
$\log\left(m_W^2/\mu^2\right)$ effects.

The renormalization group equation for the
strong coupling is
\be
\mu\frac{d}{d\mu}g_S(\mu) = \beta(g_S(\mu))
\ee
and for the Wilson coefficients
\be
\mu\frac{d}{d\mu}C_i(\mu) = \gamma_{ji}(g_S(\mu),\alpha) C_j(\mu)\,.
\ee
$\beta$ is the QCD beta function for the running coupling.

The coefficients
$\gamma_{ij}$ are the elements
of the anomalous dimension matrix $\hat\gamma$. This can be derived from
the infinite parts of loop diagrams and this has been done to
one \cite{one-loop} and two loops \cite{two-loop}.
The series in $\alpha$ and $\alpha_S$ is known to
\be
\hat\gamma = \hat\gamma^0_S \frac{\alpha_S}{4\pi} 
            + \hat\gamma^1_S \left(\frac{\alpha_S}{4\pi}\right)^2
            + \hat\gamma_e \frac{\alpha}{4\pi}
            + \hat\gamma_{se}\frac{\alpha_S}{4\pi} \frac{\alpha}{4\pi}
            +\cdots 
\ee
Some subtleties are involved in this calculation \cite{Buras1,two-loop}:\\
%\begin{itemize}
$\bullet$ 
{Definition of $\gamma_5$ is important: NDR versus 't~Hooft-Veltman}\\
$\bullet$
{Fierzing is important: how to write the operators}\\
$\bullet$
{Evanescent operators}\\

We now need to perform the following steps to get down to a scale $\mu_{OPE}$
somewhere around 1~GeV.
\begin{enumerate}
\item
Solve equations numerically or approximate analytically; run from 
$\mathbf{\mu_W}$
to $\mathbf{\mu_b}$
\item
At $\mathbf{\mu_b}\approx m_b$ remove $b$-quark and do matching to
theory without $b$
\item
run down from $\mathbf{\mu_b}$ to $\mathbf{\mu_c}\approx m_c$
\item
At $\mathbf{\mu_c}$ remove $c$-quark and do matching to
theory without $c$.
\item
run from $\mathbf{\mu_c}$ to $\mathbf{\mu_{\mbox{OPE}}}$
\end{enumerate}
This way we summed {\em all} large logarithms including $m_W$, $m_Z$, $m_t$,
$m_b$ and $m_c$. This is easily done this way, impossible
otherwise.

Notice that we had lots of scales $\mathbf{\mu_i}$.
In principle nothing depends on any of them

We now use the inputs
$m_t(m_t)=166~GeV$, $\alpha=1/137.0$, $\alpha_S(m_Z) = 0.1186$
which led to the initial conditions shown in table \ref{tabzimw}
and perform the above procedure down to $\mu_{OPE}$.
Results for 900~MeV are shown in columns two and three of Table \ref{tabzimu}.
\begin{table}
\hskip1cm
\begin{tabular}{|c|cc|cc|}
\hline
i        & $z_i$ & $y_i$ & $z_i$ & $y_i$ \\
         &{ $\mu_{OPE} = 0.9$ GeV} &{ $\mu_{OPE} = 0.9$ GeV} & 
{ $\mu_X =0.9$ GeV} & { $\mu_X =0.9$ GeV}\\
\hline
$z_{1}$ &$-$0.490           & 0.                & $-$0.788        & 0.\\
$z_{2}$	&1.266 		    & 0.                & 1.457           & 0. \\
$z_{3}$	&0.0092		    &  0.0287           & 0.0086          & 0.0399\\
$z_{4}$	&$-$0.0265	    & $-$0.0532         & $-$0.0101       & $-$0.0572\\
$z_{5}$	& 0.0065	    & 0.0018            & 0.0029          & 0.0112\\
$z_{6}$	&$-$0.0270	    & $-$0.0995         &  $-$0.0149      & $-$0.1223\\
$z_{7}$	& 2.6$~10^{-5}$	    &$-$0.9$~10^{-5}$   & 0.0002          &$-$0.00016\\
$z_{8}$	& 5.3$~10^{-5}$	    & 0.0013            &  6.8$~10^{-5}$  & 0.0018\\
$z_{9}$	& 5.3$~10^{-5}$	    & $-$0.0105         &  0.0003         & $-$0.0121\\
$z_{10}$&$-$3.6$~10^{-5}$   &0.0041             & $-$8.7$~10^{-5}$& 0.0065\\
\hline
\end{tabular}
\caption{The Wilson coefficients
$z_i$ and $y_i$ at a scale $\mu_{\mbox{OPE}}=$ 900~MeV
in the NDR scheme and in the $X$-boson scheme at $\mu_X =$ 900~MeV.}
\label{tabzimu}
\end{table}
Notice that $z_1$ and $z_2$ have changed much from $0$ and $1$.
This is the short-distance contribution to the $\Delta I=1/2$ Rule.
We also see a large enhancement of $y_6$ and $y_8$, which will
lead to our value of $\varepsilon^\prime$.

A similar exercise can be performed for $K^0$-\kob\ mixing \cite{Herrlich}.
This yields the effective Hamiltonian
\be
{\cal H}_{\mbox{eff}}^{\Delta S=2} = C_{\Delta S=2}
\left(\bar s_\alpha\gamma_\mu d_\alpha\right)_L
\left(\bar s_\beta\gamma_\mu d_\beta\right)_L
\ee
with
\be
C_{\Delta S=2} = \frac{G_F^2 M_W^2}{16\pi^2}
\left[\lambda_c^2\eta_1 S_0(x_c)
+\lambda_t^2\eta_2 S_0(x_t)+2\lambda_c\lambda_t S_0(x_c,x_t)\right]
%\nonumber\\&&\nonumber\times
\alpha_S^{(-2/9)}(\mu)
\left(1+\frac{\alpha_S(\mu)}{4\pi} J_3\right)
\ee
and
\be
x_c =\frac{m_c}{M_W^2}\quad\quad \lambda_i=-V_{id}V_{is}^*\,.
\quad \quad\eta_1 =1.53\quad\eta_2 = 0.57 \quad \eta_3=0.47
\ee
are
obtained with the same input as before.

\subsection{Step II: Matrix elements}

Now
remember that the $C_i$ depend on $\mu_{OPE}$ and on the
definition of the $Q_i$. This dependence should now cancel in the
final result.
We can solve this in various ways.
\begin{itemize}
\item {\bf Stay in QCD} $\Rightarrow$ Lattice calculations as described
in the lectures by Sachrajda. I will not comment further on these.
\item {\bf Give up} $\Rightarrow$ Naive factorization.
\item {\bf Improved factorization}
\item {\bf  $X$-boson method} (or fictitious boson method)
\item {\bf  Large $N_c$} (in combination with something like
the $X$-boson method.) Here the difference is mainly in the
treatment of the low-energy hadronic physics. Three main approaches
exist of increasing sophistication\footnote{Which of course means that
calculations exist only for simpler matrix-elements for the more sophisticated
approaches.}.
\begin{itemize}
\item CHPT: As originally proposed by Bardeen-Buras-G\'erard \cite{BBG}
and now pursued mainly by Hambye and collaborators \cite{Hambye}.
\item ENJL (or extended Nambu-Jona-Lasinio model\cite{ENJL}):
 As mainly done by myself and J.~Prades
\cite{BPBK,BPscheme,kptokpp,BPdIhalf,BPeps}.
\item LMD or lowest meson dominance approach \cite{LMD}.
\end{itemize}
\end{itemize}
Notice that there other approaches as well, e.g. the chiral quark model
\cite{Bertolini}. These have no underlying arguments why the $\mu$-dependence
should cancel.

I will no first discuss the various approaches in the framework
of $B_K$ and will later quote our results for the other quantities.

\subsubsection{Factorization and/or vacuum-insertion-approximation}

This is quite similar to the naive estimate for $K\to\pi\pi$ described
above except it is applied to the four-quark operator rather than to
pure $W$-exchange. So we use
$
\langle0|\bar s_\alpha\gamma_\mu d_\alpha|K^0\rangle =
i\sqrt{2}F_K p_K
$
to get
\be
\langle \overline K^0|{\cal H}_{eff}|K^0\rangle = C_{\Delta S=2}(\mu)
\frac{16}{3}F_K^2 m_K^2\,.
\ee
Now the other results are usually quoted in terms of this one, the ratio
is the so-called bag-parameter\footnote{Named after one of the early
models in which they were estimated.} $\hat B_K$.

So vacuum-insertion or
 factorization yields $\hat B_K\equiv 1$. Using large $N_c$,
the part where the quarks of one $K^0$ come from two different
currents in ${\cal H}_{\mbox{eff}}$ has to be excluded yields
$\hat B_K\equiv 3/4$.

\subsubsection{Improved Factorization}

This corresponds to first taking a matrix-element between a particular
quark and gluon external state of
${\cal H}_{\mbox{eff}}$. This removes the scheme and scale
dependence but introduces a dependence on the particular
quark external state chosen.
This can be found in the paper by Buras, Jamin and Weisz quoted
in \cite{two-loop} and has been extensively used by H.-Y.~Cheng\cite{cheng}.
This yields a correction factor of
\be
\left(1+r_1\frac{\alpha_S(\mu)}{\pi}\right)\quad r_1=-\frac{7}{6}
\ee
for $B_K$ and a similar correction matrix for the other cases.

\subsubsection{The $X$-boson method: a simpler case first.}

Let us look at the simpler example of the electromagnetic
contribution to
$m_{\pi^+}^2-m_{\pi^0}^2$ in the chiral limit. This contribution
comes from one-photon exchange as depicted in Fig. \ref{figpippio}.
\begin{figure}
\begin{minipage}{0.48\textwidth}
\hskip0.2cm
\SetScale{1.}
\setlength{\unitlength}{1.pt}
\begin{picture}(200,115)(0,0)
\SetWidth{1.5}
\Line(0,50)(200,50)
\Text(0,55)[lb]{$\pi^+$}
\Text(200,55)[rb]{$\pi^+$}
\PhotonArc(100,75)(30,-10,190){5}{8}
\Text(135,90)[lb]{$\gamma$}
\GOval(100,50)(30,50)(0){0.3}
\Text(100,50)[]{Strong Interaction}
\end{picture}
\caption{The electromagnetic contribution to the $\pi^+$-$\pi^0$
mass difference}
\label{figpippio}
\end{minipage}
\hskip1cm
\begin{minipage}{0.42\textwidth}
\hskip0.2cm
%\raisebox{-45pt}{
\SetScale{0.9}
\setlength{\unitlength}{0.9pt}
\begin{picture}(130,100)(-10,0)
\SetWidth{1.2}
\ArrowLine(0,90)(120,90)
\ArrowLine(120,10)(0,10)
\Text(22,50)[r]{gluon}
\Gluon(30,90)(30,10){5}{5}
\Text(95,50)[l]{$\gamma$}
\Photon(90,90)(90,10){5}{4}
\end{picture}
%}%raisebox
\caption{The short-distance photon-gluon box diagram leading to a four-quark
operator.}
\label{figboxphoton}
\end{minipage}
\end{figure}
The matrix element involves and integral over all photon momenta
\be
M=\dsp\int_0^\infty dq_\gamma^2\,.
\ee
We now split the integral at the arbitrary scale $\mu^2$.
The short-distance part of the integral,
$\dsp\int_{\mu^2}^\infty  dq_\gamma^2$,
can be evaluated
using OPE techniques \cite{BBG2} via the box diagram
of Fig. \ref{figboxphoton}. Other types of contributions are suppressed
by extra factors of $1/\mu^2$. The resulting four-quark operator
$\dsp (\bar{q}q)(\bar{q}q)$ can be estimated in large $N_c$\cite{BBG2}
\be
\left.  m_{\pi^+}^2-m_{\pi^0}^2 \right|_{\mbox{SD}} =
\frac{3\alpha_S\alpha_e}{\mu^2 F^4}\langle\bar {q}q\rangle^2\,.
\ee
The long-distance contribution,$\dsp\int_0^{\mu^2}dq_\gamma^2$,
 can be evaluated in several ways.
CHPT at order $p^2$ or $p^4$ \cite{BBG2}, a vector meson dominance model(VMD)
\cite{BBG2}, the ENJL model\cite{BPelem}
or LMD \cite{LMD}.
\begin{figure}[t]
\includegraphics[width=0.8\textwidth,angle=0]{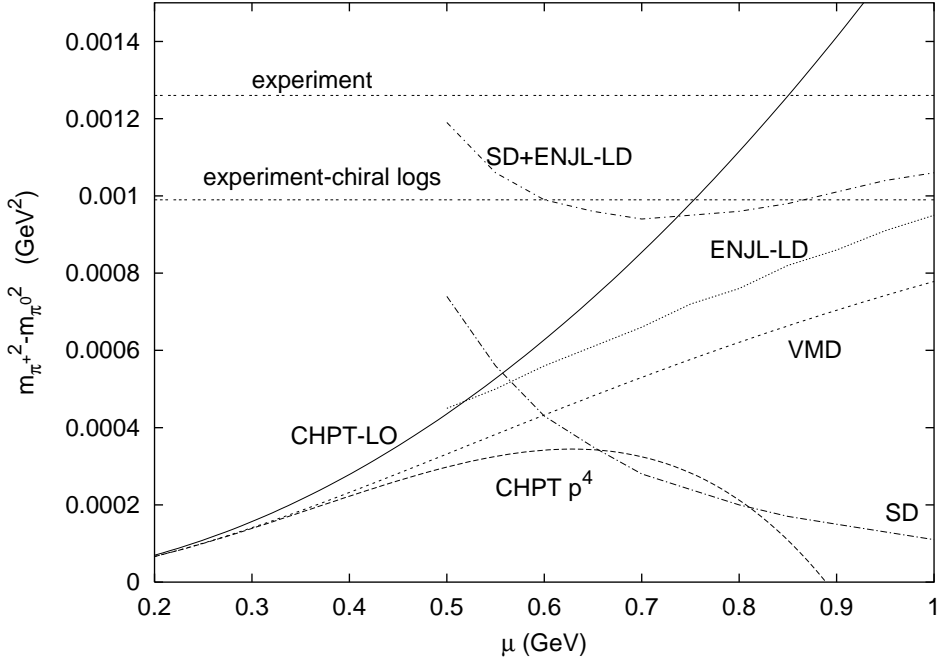}
\caption{The Short-distance contribution (SD) and the various
versions of the long-distance contributions to
$m_{\pi^2}-m_{\pi^0}^2$. Also shown are the experimental value
and the experimental value minus the chiral logarithms that are extra.}
\label{figpiresult}
\end{figure}
The results are shown in Fig. \ref{figpiresult}. Notice that the sum
of long- and short- distance contributions is quite stable in the regime
$\mu\approx 500~$MeV to 1~GeV. VMD and LMD are the same in this case.

The main comments to be remembered are:
\begin{itemize}
\item The photon couplings are known {\em everywhere}.
\item We have a good identification of the scale $\mu$. It can
be identified from the photon momentum which is unambiguous.
\item In the end we got good matching, $\mu$-independence,
 and the numbers obtained agreed (maybe too) well with the experimental
result.
\end{itemize}

\subsubsection{The $X$-boson method.}

The improved factorization model is scheme- and scale-independent but
depends on the particular choice of quark/gluon state.
Now, photons are identifiable across theory boundaries, or
more generally, currents are\footnote{At least the problem of
matching two-quark operators across theories is much more
tractable than four-quark operators.}. An example of this is CHPT
where the currents are the same as in QCD as discussed by Ecker in his
lectures.

We can now try to get our four-quark operators back into something resembling
a photon so we can use the same method as in the previous section.
The full description including all formulas can be found in \cite{BPscheme}.
For $\hat B_K$ this can be done by replacing
\be
{\cal H}_{\mbox{eff}}^{\Delta S=2}
\quad\mbox{ by }\quad
g_{X} X_{\Delta S=2}^\mu \left(\bar s_\alpha\gamma_\mu
 d_\alpha\right)_L\,.
\ee
with $M_X$ such that $\alpha_S\log\frac{M_X}{\mu}$ is small and
we can neglect higher orders in $\mu^2/M_X^2$.

\begin{figure}
\includegraphics[width=14cm]{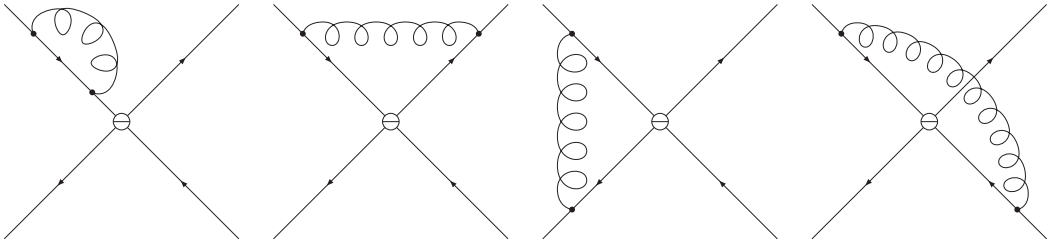}
\caption{The diagrams for the matrix-element of ${\cal H}_{\mbox{eff}}$
at one-loop.}
\label{figdiagOPE}
\end{figure}

We now take the matrix element of ${\cal H}_{\mbox{eff}}$ between
quark and gluon external states which yields from the diagrams
in Fig. \ref{figdiagOPE}
\begin{eqnarray}
&\dsp
iC_D\left[\left(1+\alpha_S(\nu)F(q_i)\right)S_1 
+\left(1+\alpha_S(\nu)F^\prime(q_i)\right)S_2\right]
&\nonumber\\&
C_D = -C(\nu)\left(1+\frac{\alpha_S(\nu)}{\pi}\left[\frac{\gamma_1}{2}
\ln\left(\frac{2q_1\cdot q_2}{\nu^2}\right)+r_1\right]\right)
&
\end{eqnarray}
with
$S_1$ and $S_2$ the tree level matrix elements between quarks of
$(\bar{s}\gamma^\mu d)_L
(\bar{s}\gamma_\mu d)_L$.

\begin{figure}
\includegraphics[width=14cm]{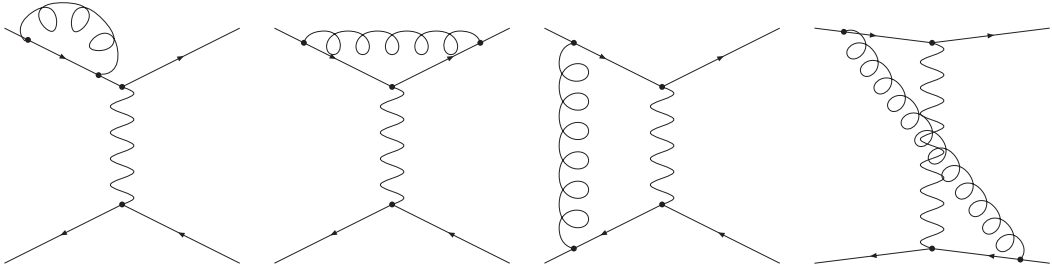}
\caption{The same matrix element but now of $X$-boson exchange.
The wiggly line is the $X$-boson.}
\label{figdiagX}
\end{figure}

We now calculate the same matrix element using $X$-boson exchange from
the diagrams in Fig. \ref{figdiagX}
and get
\begin{eqnarray}
&\dsp\hskip-1.5cm
iC_C\left[\left(1+\alpha_S(\mu_C)F(q_i)\right)S_1 +
(1+\alpha_S(\mu_C)F^\prime(q_i))S_2\right]
+{\cal O}(M_X^{-4})
\nonumber\\
&
C_C = \frac{-g_X^2}{M_X^2}
\left(1+\frac{\alpha_S(\mu_C)}{\pi}\left[\frac{\gamma_1}{2}
\ln\left(\frac{2q_1\cdot q_2}{M_X^2}\right)+\tilde{r}_1\right]\right)
\end{eqnarray}

Notice that all the dependence on the external quark/gluon state
in the functions
$F(q_i)$ and $F^\prime(q_i)$ cancels.
 $r_1$ removes the 
scheme dependence and $\tilde{r}_1$ changes to the $X$-boson current scheme.

$g_X$ is now scale, scheme and external quark-gluon state independent.
It still depends on the precise scheme used for the vector and axial-vector
current. 

The
$\Delta S=1$ case is more complicated, everything becomes
10 by 10 matrices but can be found in \cite{BPeps}.
The precise definition of the total number of $X$-bosons needed
to discuss this case is
\ba
&\dsp
g_1 X_1^\mu \left((\bar s\gamma_\mu d)_L + (\bar u\gamma_\mu u)_L\right)
+ g_2 X_2^\mu  \left((\bar s\gamma_\mu u)_L + (\bar u\gamma_\mu d)_L\right)
&\nonumber\\
&\dsp +g_3 X_3^\mu \left((\bar s\gamma_\mu d)_L + 
\sum_{q=u,d,s}(\bar q\gamma_\mu q)_L\right)
+g_4 \sum_{q=u,d,s} X_{q,4}^\mu 
    \left((\bar s\gamma_\mu q)_L+(\bar q\gamma_\mu d)_L\right)
&\nonumber\\
&\dsp
+g_5 X_5^\mu \left((\bar s\gamma_\mu d)_L + 
\sum_{q=u,d,s}(\bar q\gamma_\mu q)_R\right)
+g_6 \sum_{q=u,d,s} X_{q,6} \left((\bar s  q)_L + 
(-2)(\bar q d)_R\right)
&\nonumber\\
&\dsp
+g_7 X_7^\mu \left((\bar s\gamma_\mu d)_L + 
\sum_{q=u,d,s}\frac{3}{2}e_q(\bar q\gamma_\mu q)_R\right)
+g_8 \sum_{q=u,d,s} X_{q,8} \left((\bar s  q)_L + 
(-2)\frac{3}{2}e_q(\bar q d)_R\right)
&\nonumber\\
&\dsp
+g_9 X_9^\mu \left((\bar s\gamma_\mu d)_L + 
\sum_{q=u,d,s}\frac{3}{2}e_q(\bar q\gamma_\mu q)_L\right)
+g_{10} \sum_{q=u,d,s} X_{q,10}^\mu 
    \left((\bar s\gamma_\mu q)_L+
\frac{3}{2}e_q(\bar q\gamma_\mu d)_L\right).
&
\ea
The resulting change from this correction is displayed in columns 4 and 5
of Table \ref{tabzimu}. The corrections are substantial and turn
out to be in the wanted direction in all cases, surprisingly enough.

So let us summarize here the $X$-boson scheme
\begin{enumerate}
\item Introduce a set of fictitious gauge bosons: $X$
\item $\alpha_S\log({M_X}/{\mu})$ does not need resumming, this is not large.
\item $X$-bosons must be {\em uncolored}.
\item Only perturbative QCD and OPE have been used so far.
\item For $\hat B_K$ we need $r_1-\tilde r_1 = -\frac{11}{12}$.
\end{enumerate}

\subsubsection{$X$-boson scheme matrix element: $\hat B_K$}

We now need to calculate
$\langle \mbox{out}|X\mbox{-exchange}|\mbox{in}\rangle$. First we
do the same split in the $X$-boson momentum integral as we did
for the photon
\be
{\int d q_X^2}\quad \Longrightarrow\quad
{\int_0^{\mu^2} d q_X^2+\int_{\mu^2}^\infty d q_X^2}
\ee
For $q^2_X$ large, the
Kaon-form-factor suppresses direct mesonic contributions by $1/q_X^2$.
Large $q_X^2$ must thus flow back via {\em quarks-gluons}. The results
are already suppressed by $1/N_c$ so we can use leading $1/N_c$ in this part.
This part ends up replacing
$\dsp\log\frac{\mu_{OPE}}{M_X}$ by $\log\frac{\mu_{OPE}}{\mu}$ such that,
as it should be,
$M_X$ has disappears completely.

For the small $q^2_X$ integral we now
successively use better approximations in 3 directions:
\begin{itemize}
\item Low-energy that better approximates perturbative QCD
\item Inclusion of quark-masses
\item Inclusion of electromagnetism
\end{itemize}
The last step at present everyone only does at
 short-distance.
Chiral Symmetry provides very strong constraints, which
leads to large cancellations between various parts.

A few comments are appropriate here
\begin{itemize}
\item
The Chiral Quark Model approach \cite{Bertolini} does not do the
identification of scales and we do not include their results.
But they stressed large effects from {\em FSI, quark-masses} when
factorization+small variations was the main method. See also
\cite{PP}.
\item
For some matrix-elements CHPT allows to relate them to integrals
over measurable spectral functions, \cite{DH}.
The remainder agrees numerically for these $B_7,m_{\pi^+}^2-m_{\pi^0}^2$.
\end{itemize}

\begin{figure}
\begin{minipage}{0.47\textwidth}
\includegraphics[height=\textwidth,angle=270]{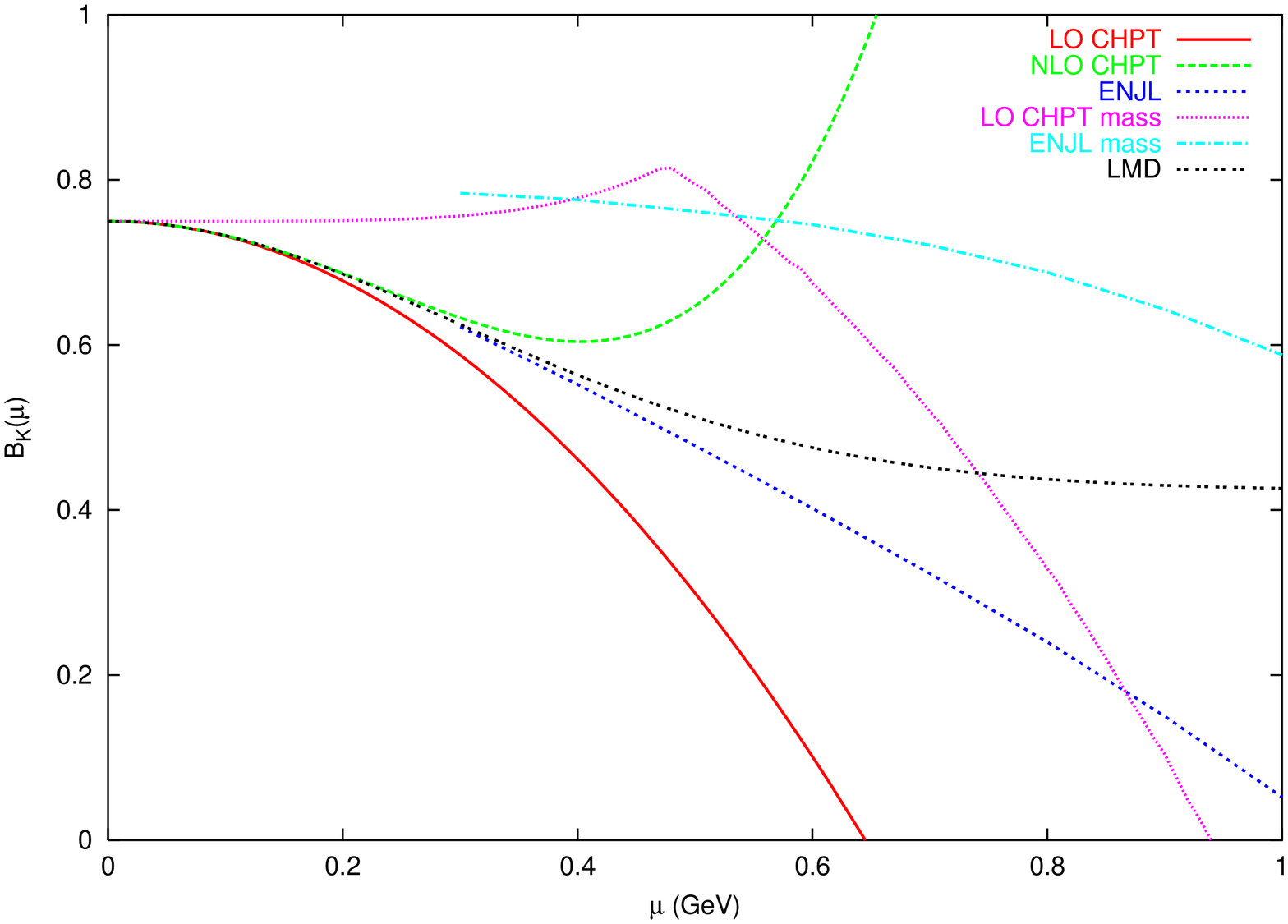}
\caption{Comparison of the long-distance contributions 
to $B_K$ in the
various approximations discussed in the text.}
\label{figBKlong}
\end{minipage}
\hskip0.03\textwidth
\begin{minipage}{0.47\textwidth}
\includegraphics[width=\textwidth]{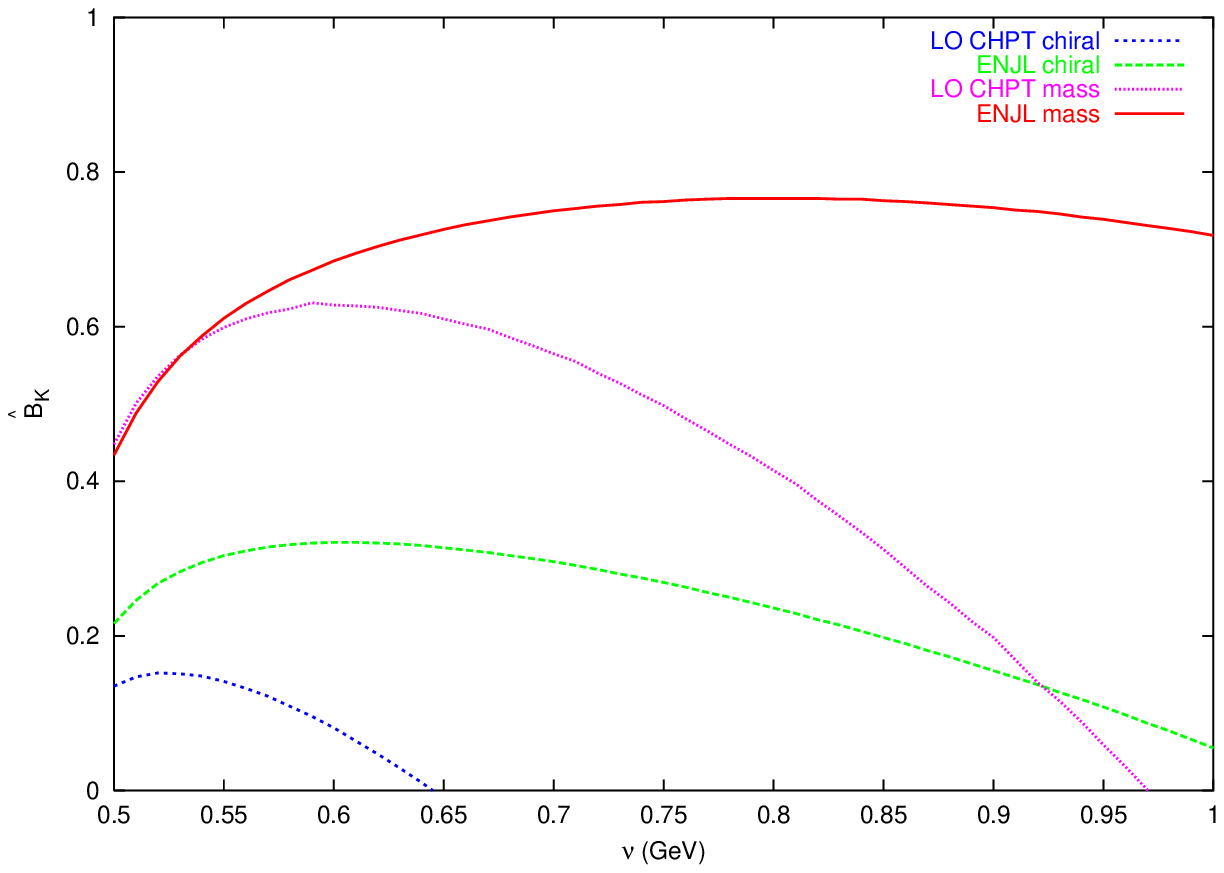}
\caption{Using the long-distance depicted in Fig. \ref{figBKlong}
we obtain as results for $\hat B_K$ as a function of $\mu$.}
\label{figBK}
\end{minipage}
\end{figure}

So the different results for
$B_K(\mu)$ in the chiral limit are
\be
B_K^\chi(\mu) =
\frac{3}{4}\left[
1\mbox{ (large-$N_c$)}
-\frac{3\mu^2}{16\pi^2F_0^2}\mbox{ ($p^2$)}
%\right.
%\nonumber\\&&
%\left.
+\frac{6\mu^4}{16\pi^2F_0^4}
\left(2L_1+5L_2+L_3+L_9\right)\mbox{ ($p^4$)}\right]
\ee
for CHPT\cite{BPBK,BPscheme}. The ENJL model
we do numerically\cite{BPBK,BPscheme}
and LMD gives\footnote{
I have pulled factors of $\mu^2_{had}$ 
into the $\alpha_i,\beta_i$}\cite{LMD}
\be
B_K^\chi(\mu) =
\frac{3}{4}\left\{1-\frac{1}{32\pi^2F_0^2}\right.
\int_0^{\mu^2} dQ^2 
%\times\nonumber\\&&
%\hskip-1.5cm
\left.\left(6-\sum_{i=res}\left[
\frac{\alpha_i}{Q^2+M_i^2}-\frac{\alpha_i}{M_i^2}
+\frac{\beta_i}{(Q^2+M_i^2)^2}-\frac{\beta_i}{M_i^4}\right]\right)\right\}
\ee
$\alpha_i$ and $\beta_i$ are particular combinations of the resonance
couplings. These we can now restrict by
comparing CHPT and LMD,
\be
\sum_i\frac{\alpha_i}{M_i^4}+\frac{\beta_i}{M_i^6}
=\frac{24}{F_0^2}\left(2L_1+5L_2+L_3+L_9\right)\,,
\ee
and using short-distance constraints:
\be
\dsp\sum_i \left(\alpha_i M_i^2 -\beta_i\right) = 0
\quad\quad
\sum_i
\left(\frac{\alpha_i}{M_i^2}+\frac{\beta_i}{M_i^4}\right)=6
\quad\quad
\sum_i \alpha_i = 24\pi^2\frac{\alpha_S}{\pi} F_0^2\,.
\ee
The last requirement is from explicitly requiring matching.
The various long distance contributions in the chiral limit and
in the presence of masses are shown in Fig. \ref{figBKlong}.
Including the short-distance part leads to the results for $\hat B_K$
shown in Fig. \ref{figBK} and
\ba
\hat B_K^\chi &=& 0.32\pm0.06~(\alpha_S)\pm0.12~(\mbox{model})\nonumber\\
\hat B_K &=& 0.77\pm0.05~(\alpha_S)\pm0.05~(\mbox{model})\,.
\ea
The LMD model leads to somewhat higher 
but compatible results for the chiral case\cite{LMD}.

\subsubsection{$X$-boson method results for $\Delta I=1/2$ rule
and $\varepsilon^\prime/\varepsilon$.}

We now present the results of the $X$-boson method also for
the $\Delta S=1$ quantities. For other approaches I refer to the
various talks given at ICHEP2000 in Osaka.
The notation used below and more extensive discussions can be found
in \cite{BPeps}.

The lowest-order CHPT lagrangian for the non-leptonic
$\Delta S=1$ sector is given by
\begin{eqnarray}
\label{CHPTdS1}
{\cal L}_{\Delta S=1} &=&
-CF_0^4\, \left[
{ G_8}\, \mbox{tr}(\Delta_{32}u_\mu u^\mu )
+ \, {G_8^\prime}\, \mbox{tr}(\Delta_{32}\chi_+ )
\right.
\nonumber \\&& 
+ \, {G_{27}} \, 
t^{ijkl}\mbox{tr}(\Delta_{ij} u_\mu u^\mu )
\mbox{tr}(\Delta_{kl}u_\mu u^\mu )
\left.+{e^2 G_E} F_0^2\mbox{tr}(\Delta_{32}\tilde Q)
\right] 
\, ; 
\end{eqnarray}
and contains four couplings,
The various notations used are
\begin{eqnarray*}
U    \equiv 
\exp\left({i\sqrt{2} \Phi/F_0}\right) \equiv u^2 \, ;\quad
u_\mu \equiv i u^\dagger  (D_\mu U) u \, ;
&&
 \chi_+   \equiv   
 2 B_0 \left ( u^\dagger {\cal M} u^\dagger + 
u {\cal M}^\dagger u \right) \, \\
\Delta_{ij}= u\lambda_{ij}u^\dagger\,;\quad\quad
(\lambda_{ij})_{ab}=\delta_{ia}\delta_{jb}\,;\quad\quad
\tilde Q = u^\dagger Q u\,;
&&
C = \frac{3G_F}{5\sqrt{2}}V_{ud}V_{us}^*
\end{eqnarray*}
and are similar to the ones used by Ecker in his lectures.

Fixing the parameters from $K\to\pi\pi$
allows to predict $K\to3\pi$ to about 30\%.

In the limit
$N_c \to \infty$ \& $e\to0$ the parameters become
\be
 G_8 = G_{27}  \to 1 \quad
 G_8'\, e^2 G_E\to 0\,. 
\ee
The isospin 0 and 2 amplitudes for $K\to\pi\pi$ from the
above Lagrangian are
\ba
{a_0} & = &\frac{\sqrt 6}{9} \, C F_0\left[\left(9G_8+G_{27}\right) 
(m_K^2-m_\pi^2) - 6 e^2 G_E F_0^2 \right]
\nonumber\\
{a_2} & = & \frac{\sqrt 3}{9} \, C F_0\left[10 G_{27} \, 
(m_K^2-m_\pi^2) - 6 e^2 G_E F_0^2 \right]\,.
\ea
The experimental values are\cite{KMW,kptokpp}
Re$(G_8)\approx 6.2$ and Re$(G_{27})\approx0.48$ with a sizable error.

\begin{figure}
\begin{minipage}[b]{0.48\textwidth}
\includegraphics[height=0.9\textwidth,angle=270]{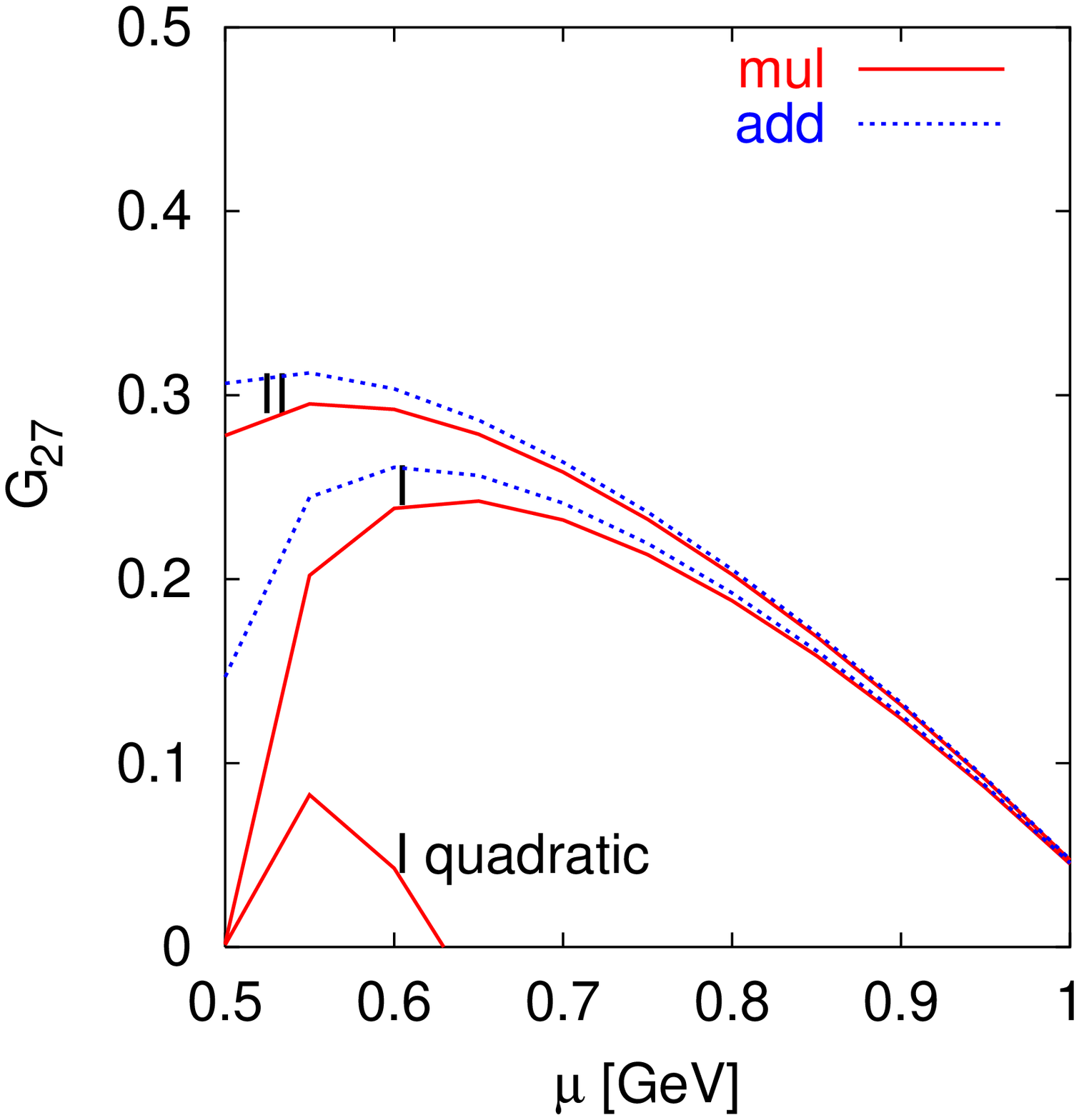}
\end{minipage}
\begin{minipage}[b]{0.48\textwidth}
\includegraphics[height=0.9\textwidth,angle=270]{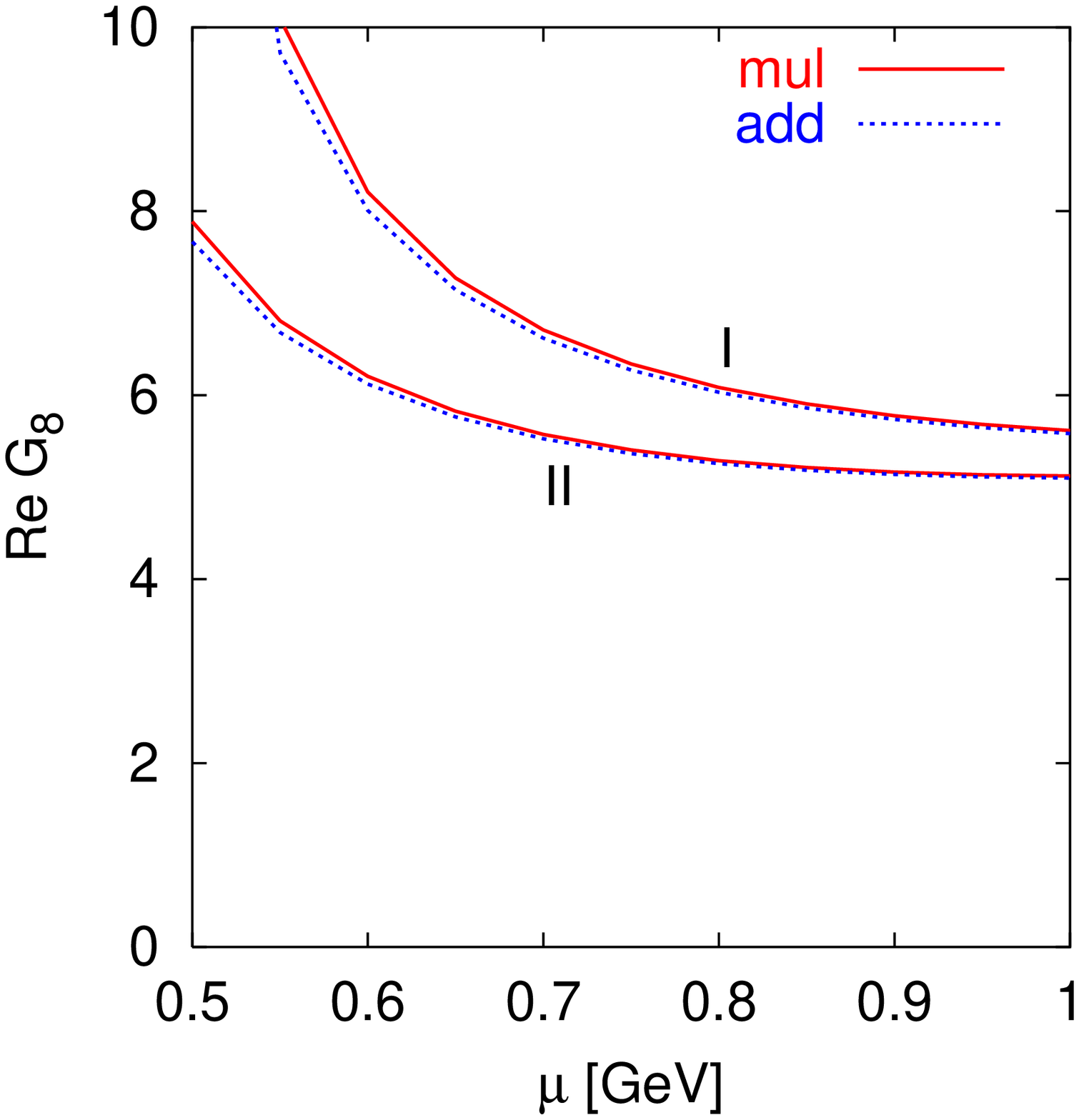}
\end{minipage}
\caption{Results for the real part of the $\Delta S=1$ chiral lagrangian.
Remember that
$\re G_8^{\mbox{exp}}\approx 6.2$ and 
$\re G_{27}^{\mbox{exp}}\approx 0.48$.
The labels I and II refer to two different values of $\alpha_S$ and
mul-add to two ways of combining the various QCD corrections, differing only
at higher orders.
}
\label{figreal} 
\end{figure}
\begin{figure}
\begin{minipage}[b]{0.48\textwidth}
\includegraphics[height=0.9\textwidth,angle=270]{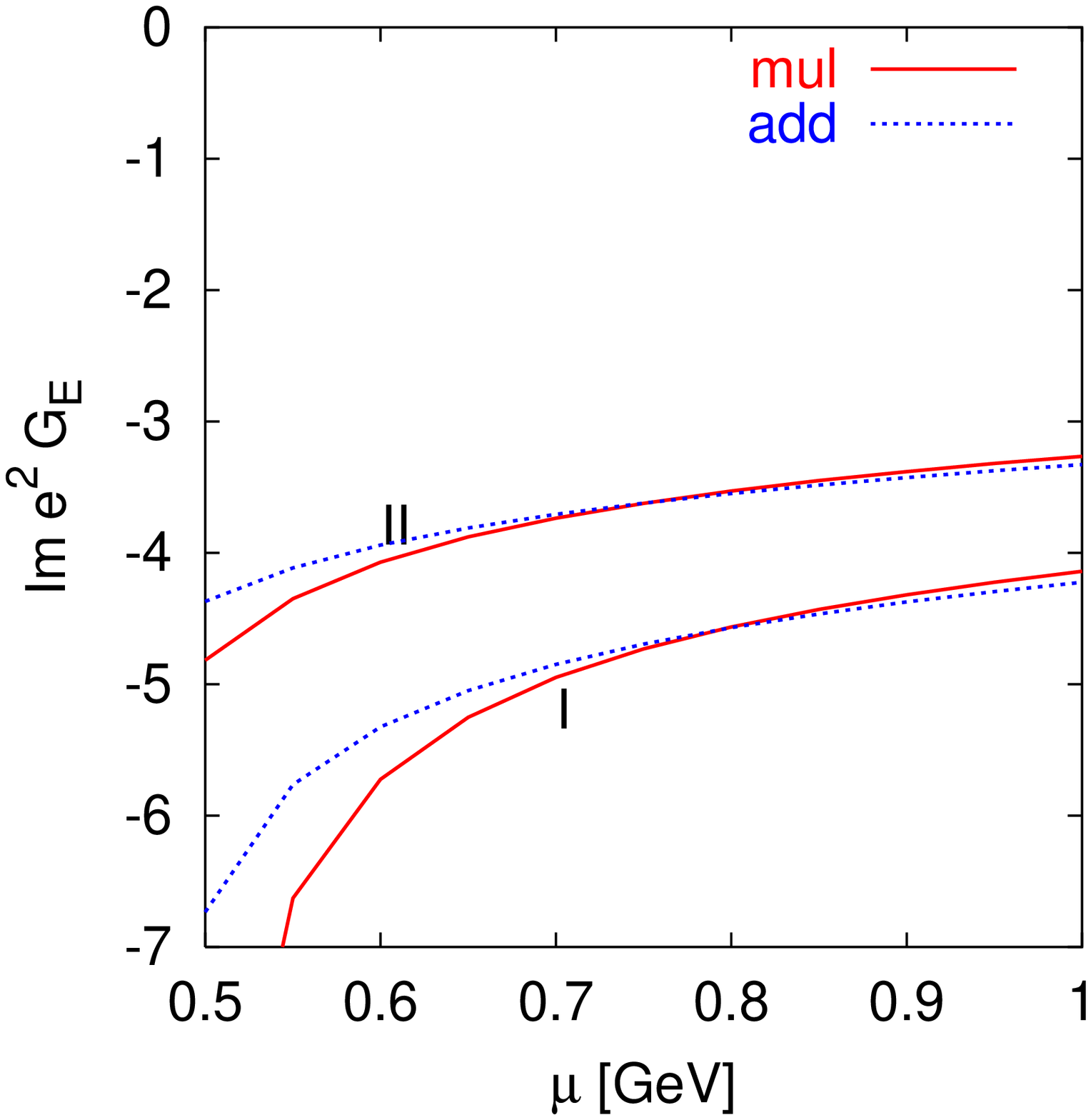}
\end{minipage}
\begin{minipage}[b]{0.48\textwidth}
\includegraphics[height=0.9\textwidth,angle=270]{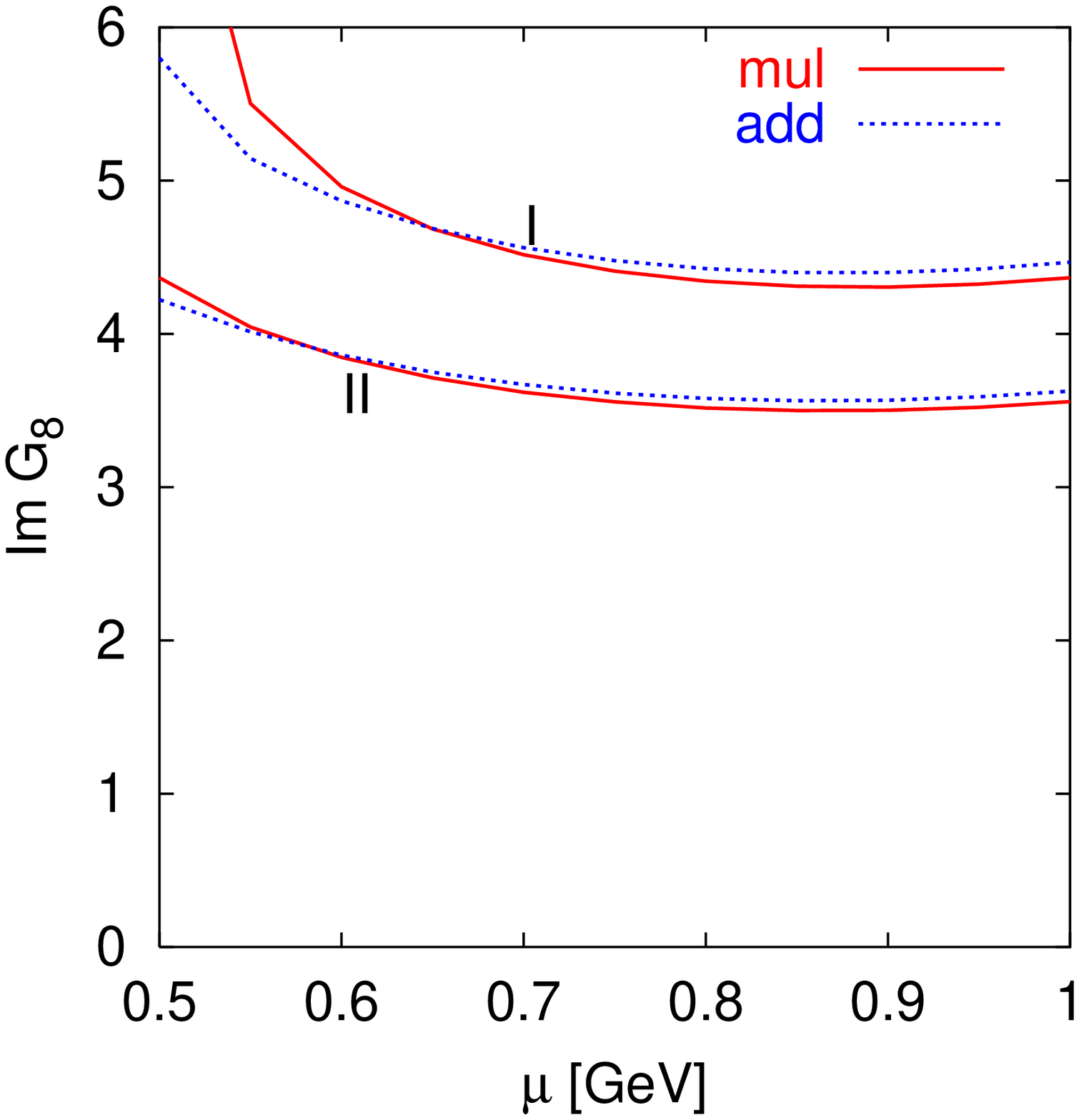}
\end{minipage}
\caption{Results for the imaginary part of the $\Delta S=1$ chiral lagrangian.}
\label{figimag}
\end{figure}

The results obtained are shown in Fig. \ref{figreal} for the real
parts and in Fig. \ref{figimag} for the imaginary parts.
Notice that we get good matching for most quantities and good agreement
with the experimental result for $G_8$. The bad matching for
$G_{27}$ is because we have a large cancellation needed
between the non-factorizable and the factorizable case to obtain matching.
The  30\% or so accuracy we have on the non-factorizable part leads therefore
to large errors on the final result. The other quantities are not affected
by such a cancellation.

We can now use these results to estimate
$\varepsilon^\prime/\varepsilon$ in the chiral limit.
We used $G_{27}=0.48$, $\re G_8=6.2$ and the values we obtained for the
imaginary part. The same method leads to
$\varepsilon$ within 10\% of the experimental value. The result
is shown in Fig. \ref{figeps}.
\begin{floatingfigure}{8cm}
\includegraphics[height=7.5cm,angle=270]{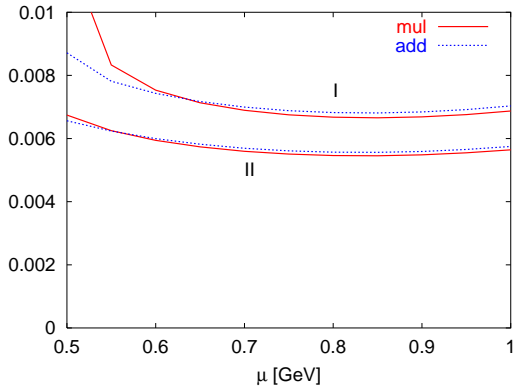}
\caption{The results for $\varepsilon^\prime/\varepsilon$ in the chiral limit.
Notice the quite good matching.}
\label{figeps}
\end{floatingfigure}
We can conclude that
\be
\dsp\left(\frac{\varepsilon^\prime}{\varepsilon}\right)^\chi
= (7.4-1.9)\cdot10^{-3}=5.5\cdot10^{-3}
\ee
and\\
$\bullet$
 $B_6\approx 2.5$ not $\lesssim 1.5$\\
$\bullet$  $B_8\approx 1.3$ OK but not $B_8\approx B_6$\,.

Using
\be
\label{epsp}
|\varepsilon^\prime| \simeq
\frac{1}{\sqrt{2}}\frac{\re \, a_2}{\re \, a_0}
\left(-\frac{\im \, a_0}{\re \, a_0}+\frac{\im \, a_2}{\re \, a_2}\right)
\ee
We can now include the two main known corrections.
The usual approach for final state interactions (FSI)
is to take
$\re a_0$, $\re a_2$ from experiment and
$\im a_0$, $\im a_2$ to ${\cal O}(p^2)$. This leads
to a large suppression of the first term in (\ref{epsp})\cite{PP}.
We evaluate both to $p^2$ so for us
FSI act mainly on the  prefactor in Eq. (\ref{epsp}).

The main isospin breaking correction
is that
$\pi^0$,$\eta$ and $\eta^\prime$. This brings in a part
of the large $a_0$ into $a_2$ and is thus enhanced. The
effect is usually parametrized as
\be
\frac{\Delta\im a_2}{\re a_2}\approx\Omega\frac{\im a_0}{\re a_0}
\quad\mbox{with}\quad \Omega\approx 0.16\pm0.03
\ee
where the numerical value is taken from \cite{EPetal}.

Including the last two main corrections yields
\be
\left|\frac{\varepsilon^\prime}{\varepsilon}\right| = (5.4-2.3)\cdot10^{-3}
= (3.1\pm ??)\cdot10^{-3}\,.
\ee
The size of the error is debatable but should be at least 50\% given
all the uncertainties involved.

\section{CHPT tests in non-leptonic Kaon decays to pions}

I have already shown the lowest order $\Delta S=1$ CHPT lagrangian
in Eq. (\ref{CHPTdS1}) and mentioned that this
reproduces $K\to\pi\pi\pi$ to about 30\% from $K\to\pi\pi$.
This can now be extended to an order $p^4$ calculation in CHPT\cite{KMW,KMW2}.
%The diagrams needed, now the lines are mesons, not quarks as in most
%of the previous figures, are shown in fig \ref{figk2pi} for $K\to\pi\pi$.
%The $K\to\pi\pi\pi$ ones are similar.
%\begin{figure}
%\includegraphics[width=0.9\textwidth]{kaon99_weak.ps}\\
%\caption{The CHPT diagrams with meson loops at order $p^4$
%for $K\to\pi\pi$.}
%\label{figk2pi}
%\end{figure}
In terms of the number of parameters and observables we
have
\begin{tabbing}
\# parameters :\= $p^2$ :\= 2 {(1)} \= $G_8$, $G_{27}$\\
\> $p^4$ :\> 7{(3)} \>
($-2$ that cannot be disentangled \\ \>\>\> from $G_8$, $G_{27}$)\\[0.5cm]
\# observables:\> After isospin\\[0.2cm]
$K\to 2\pi$ :\> 2{(1)}\\
$K\to3\pi$ :\> 2{(1)}{+1}\>\>constant in Dalitz plot\\
\> 3{(1)}{+3}\>\>linear\\
\> 5{(1)}\>\>quadratic\\
\end{tabbing}
Phases (the {+i} above) of up to linear terms in the
Dalitz plot might also be measurable.
The numbers in brackets refer to $\Delta I=1/2$ parameters and observables
only. Notice that a significant number of tests is possible.
Comparison with the present data is shown in Table \ref{tabk3pi}.
The numbers in brackets refer to which inputs produce which predictions.
\begin{table}
\hskip3cm
\begin{tabular}{|c|ccc|}
\hline
\mbox{variable} & $p^2$ & $p^4$ &\mbox{experiment} \\
\hline
$\alpha_1 $& 74 & (1)\mbox{input} & 91.71$\pm$0.32\\
$\beta_1  $& $-$16.5  & (2)\mbox{input} & $-$25.68$\pm$0.27\\
$\zeta_1  $& & (1)$-$0.47$\pm$0.18 & $-$0.47$\pm$0.15\\
$\xi_1    $& & (2) $-$1.58$\pm$0.19 & $-$1.51$\pm$0.30\\
$\alpha_3 $& $-$4.1 &(3) \mbox{input} & $-$7.36$\pm$0.47\\
$\beta_3  $& $-$1.0 &(4) \mbox{input} & $-$2.42$\pm$0.41\\
$\gamma_3 $& 1.8  &(5) \mbox{input} & 2.26$\pm$0.23\\
$\xi_3    $& 6 & (4) 0.92$\pm$0.030 & $-$0.12$\pm$0.17\\
$\xi_3^\prime$& &(5) $-$0.033$\pm$0.077 & $-$0.21$\pm$0.51\\
$\zeta_3  $& &(3) $-$0.0011$\pm$0.006 & $-$0.21$\pm$0.08\\
\hline
\end{tabular}
\caption{CHPT to order $p^4$ for $K\to\pi\pi\pi$. The variables
refer to various measurables in the Dalitz plot.
$K\to\pi\pi$ is always used as input. Numbers in brackets
indicate relations.}
\label{tabk3pi}
\end{table}
It is important that in the future experiments tests these relations directly.
At present there is satisfactory agreement with the data.
Notice that new CPLEAR data decrease the errors somewhat.

CP-violation in $K\to3\pi$ will be very difficult. The strong phases
 needed to interfere are just
too small (\cite{KMW2} last reference).
E.g. 
$\delta_2-\delta_1$ in $K_L\to\pi^+\pi^-\pi^0$ is expected to be $-0.083$
and present experiment. is only $-0.33\pm0.29$.
The CP-asymmetries expected are about $10^{-6}$
so we expect in the near future only to improve limits.

\section{Kaon rare decays}

The below is a summary of the summary by Isidori given at
KAON99\cite{Isidori}. I refer there for references. Another somewhat
older but more extensive review is \cite{Littenberg} and I also
found \cite{Buras2} useful.

Some of the processes mentioned below are tests of strong interaction physics,
often in the guise of CHPT, and others are mainly SM tests.

\begin{itemize}
\item
{$\mathbf K^+\to\pi^+\nu\bar\nu$,$ \mathbf K_L\to\pi^0\nu\bar\nu$}
In this case the SM is strongly suppressed and dominated by
short-distance physics. It is thus
ideal for precision SM CKM tests and
possibly new physics searches.
The reason is that real and imaginary part of the amplitude are
similar here in size, CKM angle suppression is counteracted by the
large top-quark mass. This allows it to be dominated 
by $\bar s d Z$-Penguin and $WW$-box diagrams.
The resulting
\be
{\cal H}_{\mbox{eff}} = C_\nu (\bar s \gamma_\mu d)_L(\bar\nu\gamma^\mu\nu)_L
\ee
can be hadronized using the measured matrix-element from $K_{\ell3}$
and 
$\bar\nu\nu$ is in a CP eigenstate allowing lots of CP-tests.
The main disadvantage is the extremely low predicted branching ratio
of
\ba
\mbox{Neutral mode: }&& (3.1\pm1.3)\cdot10^{-11}\nonumber\\
\mbox{Charged mode: }&& (8.2\pm3.2)\cdot10^{-11}
\ea
This process will be competitive with $B$-decays in next generation of
Kaon experiments.
\item
{$\mathbf K_L\to\ell^+\ell^-$}: The short-distance contribution comes
from $Z$-penguin and boxes. The main uncertainty comes from the
long-distance 2$\gamma$ intermediate state.

$K_L\to \mu^+\mu^-$ dominated by unitary part of $K_L\to\gamma\gamma$,
which can be taken from the branching ratio for that
decay. It fits the data well.

The long distance part of
$K_L\to e^+ e^-$ is more dependent on the contributions with off-shell
photons. Here there is still work to do.
\item {$\mathbf K\to\pi\ell^+\ell^-$}
The
real parts can be predicted by CHPT at order $p^4$ from 2 parameters, it
fits well.
For the imaginary part there are 
problems with long-distance contributions from $K\to\pi\gamma\gamma$.
but the CP-violating quantities are often dominated by direct part.
\item {$\mathbf K_S\to\gamma\gamma$}
This process was a parameter-free prediction from CHPT
 at order $p^4$ from the diagrams in Fig. \ref{figkgg}
\begin{figure}
\begin{minipage}{0.58\textwidth}
\setlength{\unitlength}{1.8pt}
\SetScale{1.8}
\SetWidth{0.75}
\begin{picture}(40,35)(-10,-10)
\Line(0,0)(10,0)
\Text(0,0)[r]{$K_S$ }
\CBoxc(10,0)(3,3){Black}{Black}
\Photon(10,0)(30,00){1}{4}
\Photon(10,0)(30,-10){1}{4}
\CArc(10,10)(10,0,360)
\Text(10,23)[b]{$\pi^+,K^+$}
\end{picture}
\setlength{\unitlength}{1.8pt}
\SetScale{1.8}
\SetWidth{0.75}
\begin{picture}(40,20)(-5,0)
\Line(0,0)(10,0)
\Text(0,0)[r]{$K_S$ }
\CBoxc(10,0)(3,3){Black}{Black}
\Photon(10,0)(30,00){1}{4}
\Photon(10,20)(30,20){1}{4}
\Vertex(10,20){2}
\CArc(10,10)(10,0,360)
\Text(10,23)[b]{$\pi^+,K^+$}
\end{picture}
\setlength{\unitlength}{1.8pt}
\SetScale{1.8}
\SetWidth{0.75}
\begin{picture}(40,20)(0,-10)
\Line(0,0)(10,0)
\Text(0,0)[r]{$K_S$ }
\CBoxc(10,0)(3,3){Black}{Black}
\Photon(27.07,7.07)(47.7,7.07){1}{4}
\Photon(27.07,-7.07)(47.7,-7.07){1}{4}
\Vertex(27.707,7.07){2}
\Vertex(27.707,-7.07){2}
\CArc(20,0)(10,0,360)
\Text(20,13)[b]{$\pi^+,K^+$}
\end{picture}
\caption{The meson-loop diagrams contributing to $K_S\to\gamma\gamma$.
They predict the rate well.}
\label{figkgg}
\end{minipage}
\hskip0.02\textwidth
\begin{minipage}{0.38\textwidth}
\setlength{\unitlength}{2pt}
\SetScale{2}
\begin{picture}(100,20)(0,-10)
\SetWidth{0.75}
\Line(0,0)(20,0)
\Text(10,3)[br]{$K_L$}
\CBoxc(20,0)(4,4){Black}{Black}
\Line(20,0)(60,0)
\Text(40,3)[b]{$\pi^0,\eta,\eta^\prime$}
\Vertex(60,0){2}
\Photon(60,0)(80,10){1}{4}
\Photon(60,0)(80,-10){1}{4}
\end{picture}
\caption{The main diagram for $K_L\to\gamma\gamma$ with a large
uncertainty due to cancellations.}
\label{figklgg}
\end{minipage}
\end{figure}
\item {$\mathbf K_L\to\gamma\gamma$}
This decay needs more work. The underlying difficulty is that
the main contribution is full of cancellations. The main diagram
is shown in Fig. \ref{figklgg}
\item {$\mathbf K_L\to\pi^0\gamma\gamma$}
This process at $p^4$ is again a parameter-free CHPT prediction.
The spectrum is well described but the rate is somewhat off.
This can be explained by $p^6$ effects.
\item {$\mathbf K_S\to\pi^0\gamma\gamma$}
This process has very similar problems as in $K_L\to\gamma\gamma$
\item {$\mathbf K_{L(S)}\to\gamma^*\gamma^*$}
The same processes as above but with one or both photons off-shell, decaying
into a $\ell^+\ell^-$-pair.
These have similar questions/problems/successes as the ones with on-shell
photons.

\end{itemize}

\section{Conclusions} 

\begin{itemize}
\item {\bf Semi-leptonic Decays}
\begin{itemize}
\item CHPT is a major success and tool here.
\item These decays are the main input for $V_{ud}$ and $V_{us}$
\item In addition they provide several tests of strong interaction effects.
\end{itemize}

\item {\bf $\mathbf K\to\pi\pi$ and $\mathbf \kob$-$\mathbf K^0$ mixing}
This was the main part of the lectures.
I hope I have convinced you that successful prediction is possible but more
work on including extra effects and pushing down the uncertainty is 
obviously needed.
\item{$\mathbf K\to\pi\pi\pi$} A good test of CHPT
\item{\bf Rare Decays}. I only presented a very short
summary of the issues.
\end{itemize}

\section*{Acknowledgements}
I would like to thank the organizers for a most enjoyable atmosphere
in and around the lectures. I certainly enjoyed giving these lectures.
I hope the students had similar feelings about receiving them.
This work has been partially supported  by the Swedish Science Foundation.
and by the European Union TMR Network
EURODAPHNE (Contract No. ERBFMX-CT98-0169).

\end{document}